\documentclass{aa}

\usepackage{graphicx}
\usepackage{amsbsy}
\usepackage{txfonts}
\usepackage[pdftitle={Formation of hot Jupiters through disk migration and evolving stellar tides}, colorlinks = true, breaklinks = true, citecolor = blue, linkcolor = blue, urlcolor = blue, pdfauthor = {Ren\'{e} Heller}]{hyperref}

\begin{document}

\title{Formation of hot Jupiters through\\ disk migration and evolving stellar tides}
%\subtitle{I. Overviewing the $\kappa$-mechanism}

\author{Ren\'{e} Heller\inst{1}}

\institute{Max Planck Institute for Solar System Research, Justus-von-Liebig-Weg 3, 37077 G\"ottingen, Germany\\ \href{mailto:heller@mps.mpg.de}{heller@mps.mpg.de}}

\date{Received 24 May 2018; Accepted 30 June 2019}

% \abstract{}{}{}{}{} 
% 5 {} token are mandatory
 
\abstract{Since the discovery of Jupiter-sized planets in extremely close orbits around Sun-like stars, several mechanisms have been proposed to produce these ``hot Jupiters''. Here we address their pile-up at 0.05 AU observed in stellar radial velocity surveys, their long-term orbital stability in the presence of stellar tides, and their occurrence rate of 1.2\,$\pm$0.38\,\% in one framework. We calculate the combined torques on the planet from the stellar dynamical tide and from the protoplanetary disk in the type-II migration regime. The disk is modeled as a 2D nonisothermal viscous disk parameterized to reproduce the minimum-mass solar nebula. We simulate an inner disk cavity at various radial positions near the star and simulate stellar rotation periods according to observations of young star clusters. The planet is on a circular orbit in the disk midplane and in the equatorial plane of the star. We show that the two torques can add up to zero beyond the corotation radius around young, solar-type stars and stop inward migration. Monte Carlo simulations with plausible variations of our nominal parameterization of the star-disk-planet model predict hot-Jupiter survival rates between about 3\,\% (for an $\alpha$ disk viscosity of $10^{-1}$) and 15\,\% (for $\alpha=10^{-3}$) against consumption by the star. Once the protoplanetary disk has been fully accreted, the surviving hot Jupiters are pushed outward from their tidal migration barrier and pile up at about 0.05\,AU, as we demonstrate using a numerical implementation of a stellar dynamical tide model coupled with stellar evolution tracks. Orbital decay is negligible on a one-billion-year timescale due to the contraction of highly dissipative convective envelopes in young Sun-like stars. We find that the higher pile-up efficiency around metal-rich stars can at least partly explain the observed positive correlation between stellar metallicity and hot-Jupiter occurrence rate. Combined with the observed hot-Jupiter occurrence rate, our results for the survival rate imply that ${\lesssim}\,8\,\% \, (\alpha=10^{-3})$ to ${\lesssim}\,43\,\% \, (\alpha=10^{-1})$ of sun-like stars initially encounter an inwardly migrating hot Jupiter. Our scenario reconciles models and observations of young spinning stars with the observed hot-Jupiter pile up and hot-Jupiter occurrence rates.}


\maketitle

\section{Introduction}
\label{sec:introduction}

Soon after the surprising detection of Jupiter-mass planets in very close orbits around Sun-like stars \citep{1995Natur.378..355M}, it was proposed that these hot Jupiters cannot have formed in situ but that they must have migrated from the cold, icy regions of the protoplanetary disk at several astronomical units (AU) from the star \citep{1996Natur.380..606L}. Competing theories have been put forward as to what stops the inward migration of planets \citep[for a recent review see][]{2018ARA&A..56..175D}: tidal halting \citep{1998ApJ...500..428T}, magnetorotational instabilities that evacuate the close-in protoplanetary disk \citep{2002ApJ...574L..87K,2006ApJ...645L..73R}, planet-disk magnetic interactions \citep{2003MNRAS.341.1157T}, the Kozai mechanism of a distant perturber \citep{2008ApJ...678..498N}, planet traps \citep{Hasegawa2010}, planet-planet scattering \citep{Naoz2011,Wu2011}, and high-eccentricity migration \citep{Wang2017}.

Although the tidal stopping mechanism offers the best agreement with observations \citep{2013ApJ...769...86P}, none of the previous theories for hot Jupiter formation could explain the following observations simultaneously: (1) the sharp pile-up of hot Jupiters at 0.05\,AU around Sun-like stars observed in radial velocity (RV) surveys; (2) the increase of the hot-Jupiter occurrence rate with stellar metallicity in the {\it Kepler} planet sample \citep{2018AJ....155...89P}; and (3) the long-term orbital stability of hot Jupiters under the effect of tidal dissipation in the star. These are the points that we address in this study.

At $\lesssim0.1$\,AU, tidal dissipation in the star is sufficiently large to affect the planetary orbit. Planets on circular orbits with an orbital plane near the stellar equatorial plane and with orbital semi-major axes ($a$) larger than the stellar corotation radius ($r_{\rm co}$) are repelled, whereas planets interior to $r_{\rm co}$ are driven into an ever faster orbital decay until they are either tidally disrupted or they fall into the star. Most previous studies used equilibrium tide models with an assumed fixed tidal dissipation constant \citep[$Q_\star$, typically chosen between $10^5$ and $10^6$;][]{1996Natur.380..606L,1998ApJ...500..428T,2004ApJ...610..464D,2012MNRAS.425.2567R} to parameterize tidal dissipation in the star and to evaluate the tidally driven orbital circularization and migration of close-in planets. The resulting tidal torque is too weak to stop a migrating Jupiter-mass planet.

Moreover, constant-$Q_\star$ \citep[or rather constant-angle or constant-phase-lag;][]{2009ApJ...698L..42G} models with a host star that has a Sun-like rotation period predict a gradual infall of hot Jupiters into their stars on a timescale of one billion years. Although the equilibrium tide model is compatible with the low frequency of planets within 0.03\,AU around Sun-like stars, it requires a delicate fine-tuning of the constant stellar dissipation factor or of the initial conditions in the protoplanetary disk \citep{2012MNRAS.425.2567R} to explain the existence and even pile-up of hundreds of known hot Jupiters at about 0.05\,AU around Sun-like stars.

%**********************************************
%Fig. 1
\begin{figure}
\centering
\includegraphics[angle= 0, width=1\linewidth]{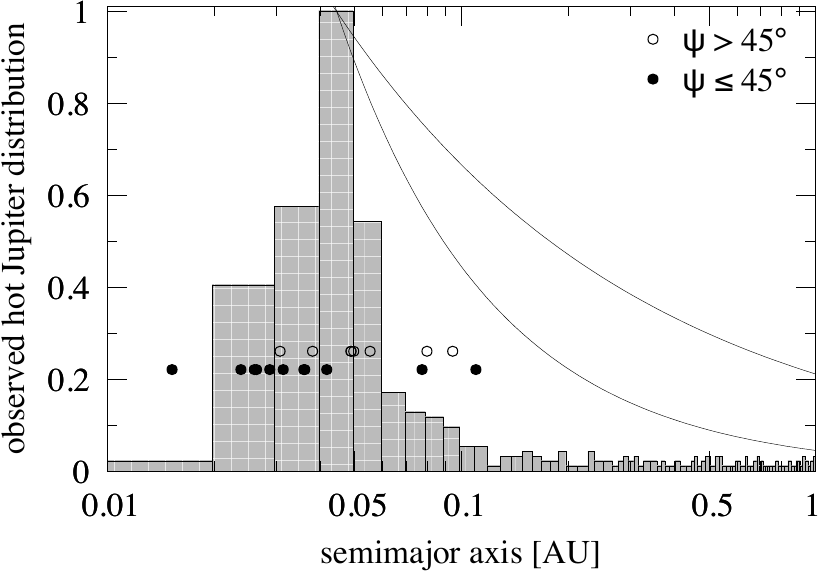}
\caption{Normalized histogram of the observed semimajor axis distribution of all extrasolar planets with known semimajor axes and masses $>0.1\,M_{\rm J}$ around stars with masses $0.75\,M_\odot~{\leq}~M_\star~{\leq}~1.25\,M_\odot$ (672 in total). Data are from the Extrasolar Planets Encyclopaedia at \href{http://exoplanet.eu}{http://exoplanet.eu} \citep{2011A&A...532A..79S} as of 9 January 2019. The solid lines illustrate the $1/\sqrt{a}$ and $1/a$ dependences of the RV amplitude and the geometric transit probability, respectively. Orbital positions of 19 planets with known stellar obliquities are shown with open ($\psi>45^\circ$) and closed ($\psi\leq45^\circ$) circles. While the scaling of the abscissa is logarithmic, the bin width is constant to suppress binning artefacts (see Appendix~\ref{sec:appendix_pileup}).}
\label{fig:exoplanets}
\end{figure}
%**********************************************

The efficiency of tidal dissipation in the star is determined by the presence and extent of the convective envelope of the star \citep{1977A&A....57..383Z,2007ApJ...661.1180O}. While Sun-like stars on the main sequence have their core--envelope boundary at about 0.7 solar radii ($R_\odot$), pre-main sequence stars are much larger than our Sun today and can have much more extended envelopes. As a consequence, while solar-type stars on the main sequence respond to the tidal perturbation by close-in massive planets with a tidal dissipation function of $10^5~{\lesssim}~Q_\star~{\lesssim}~10^8$, young stars are much more dissipative \citep{2016CeMDA.126..275B} with $Q_\star$ as low as $\approx10^3$, depending on the exact structure of a given star. The source of this dissipation is in the so-called dynamical tide within the  convective regions of the star, which is due to inertial waves that are caused by the Coriolis acceleration \citep{2007ApJ...661.1180O}. Inertial waves are driven as long as the modulus of the orbital mean motion of the planet is $|n|<2|\Omega_\star|$, where $\Omega_\star$ is the stellar spin rate. As the tidal dissipation, and therefore the tidally induced orbital decay of the planet, depends on both $n$ and $\Omega_\star$, a consistent picture for the tidal migration of hot Jupiters requires a model of the stellar spin evolution and its effects on the transfer of rotation to orbital angular momentum. Toward the end of the lifetime of a star, stellar mass loss and the engulfment of the planet within the extended gaseous envelope of the star can further affect the orbital evolution of a planet under the effect of the dynamical tide \citep{2018A&A...618A..18R}.

With $n$ and $\Omega_\star$ affecting the tidal dissipation regime, the initial rotational spin-up and subsequent magnetic braking of a star are key ingredients to hot-Jupiter formation and evolution. \citet{1999AJ....117.2941S} found that solar-mass stars in the $\sim$1\,Myr old Orion Nebula have typical rotation periods ($P_{\rm rot}$) between about 0.5 and 8\,d (the upper threshold being uncertain due to observational biases), implying corotation radii between about 0.01 and 0.08\,AU. The rotation periods of stars with ages between 1\, and 5\,Myr cluster between breakup periods ($\sim$0.5\,d) and about 8\,d with a long tail in the period distribution up to $\sim$30\,d \citep{2008MNRAS.384..675I}. We use this information to parameterize our disk model and the resulting torques on the planet (Sect.~\ref{sec:disk}).

Figure~\ref{fig:exoplanets}(a) shows the well-known pile-up in the semimajor axis distribution of all known exoplanets with masses $>0.1$ Jupiter masses ($M_{\rm J}$) around sun-like stars. The open and closed circles refer to hot Jupiters with known stellar spin-orbit obliquities ($\psi$), which illustrates that the pile up is present in both low- and high-obliquity hot Jupiters.\footnote{Data from the TEPCAT catalog at \href{http://www.astro.keele.ac.uk/jkt/tepcat/}{www.astro.keele.ac.uk/jkt/tepcat/} \citep{2011MNRAS.417.2166S}.} Appendix~\ref{sec:appendix_pileup} shows the same data plotted along a linear abscissa. In this paper, we develop a theory of hot-Jupiter formation that can at least partly explain the existence of a pile-up in the sample of {\it Kepler} planets with RV measurements.

\section{Methods}
\label{sec:methods}

In the early phase of planet formation, giant planets supposedly form beyond the circumstellar ice line at a few AU around Sun-like stars \citep{1981PThPS..70...35H}. They then migrate to close-in orbits at about 0.1\,AU or less within the protoplanetary disk. Before a protoplanet has accreted sufficient mass to open up a gap in the disk, its radial drift is referred to as type-I migration and it is driven by the Lindblad torque ($\Gamma_{\rm LB}$) and, in certain cases, the corotation torque \citep{1979ApJ...233..857G,1986ApJ...309..846L}. Planets with masses similar to or larger than that of Jupiter however open up a gap in the disk, which then leads to type-II migration on the viscous timescale of the disk \citep{1997Icar..126..261W,2000MNRAS.318...18N}\footnote{But see \citet{2014ApJ...792L..10D} and \citet{2015A&A...574A..52D} whose simulations suggest that gap opening does not necessarily couple planetary migration to the evolution of the viscous disk.}. As we are interested in hot Jupiters in this study, we consider disk torques on the planet in the type-II migration regime ($\Gamma_{\rm II}$).

If the tidal dissipation in the star is strong enough, the inward migration of a planet may halt at a stellar distance where the torque on the planet exerted by stellar tide ($\Gamma_{\rm t}$) compensates for $\Gamma_{\rm II}$, that is, where $\Gamma_{\rm t}+\Gamma_{\rm II}=0$. We refer to this distance as the tidal migration barrier.

\subsection{Disk model}
\label{sec:disk}

The disk torque on the planet depends on the local disk properties. We assume that the planet orbits the star in the disk midplane, which has a temperature $T_{\rm m}$. We model a rotationally symmetric, 2D optically gray disk with a vertical temperature gradient determined by the viscous heating of the disk and by the stellar irradiation. The disk effective temperature is given by \citep{1990ApJ...351..632H}

\begin{equation}\label{eq:t_effd}
T_{\rm eff,d} = \frac{4}{3} \ \frac{ T_{\rm m}^4 - T_{\rm i}^4 }{\tau_{\rm ext}/2 + 1/\sqrt{3} +1/(3\tau_{\rm abs})} \ ,
\end{equation}

\noindent
where $T_{\rm i}$ is the temperature due to stellar illumination, $\tau=\kappa \Sigma_{\rm p}/2$ with $\kappa_{\rm ext}$ as the Rosseland mean extinction opacity, $\kappa_{\rm abs}$ as the Rosseland mean absorption opacity, and $\Sigma_{\rm p}$ as the disk gas surface density at the position of the planet. The use of mean Rosseland mean extinction opacities is justified because the Planck mean and Rosseland mean extinction opacities are comparable in fully mixed dusty disks \citep{1994ApJ...421..615P}. Furthermore, following \citet{1994ApJ...421..615P}, we set $\kappa_{\rm abs}=\kappa=\kappa_{\rm ext}$ because it has been shown that this is adequate for disk temperatures and optical depths around accreting stars \citep{2004ApJ...606..520M}. Hence, $\tau_{\rm abs}=\tau=\tau_{\rm ext}$.

We consider a two-faced disk in thermodynamic equilibrium so that its cooling rate $Q^- = 2\sigma_{\rm SB}T_{\rm eff,d}^4$ is equal to its heating rate $Q^+=\frac{9}{4}\nu\Sigma_{\rm p}\Omega^2$ \citep{2004ApJ...606..520M}. We assume that viscous heating is by far dominant within 0.2\,AU of the star ($T_{\rm i}~{\ll}~T_{\rm m}$), the regime we are interested in. We make use of the \citet{1973A&A....24..337S} relation that describes the disk viscosity as $\nu = \alpha c_{\rm s}^2/n$ with $\alpha$ as the kinematic viscous efficiency parameter of the disk and $n = (G[M_{\rm p}+M_\star]/a^3)^{1/2}$ as the Keplerian orbital frequency at the orbital radius $a$. With the speed of sound given as $c_{\rm s} = \sqrt{ k_{\rm B} T/\mu}$, where $k_{\rm B}$ is the Boltzmann constant and $\mu$ is the mean molecular weight of the gas in units of the proton mass ($m_{\rm p^+}$), we transform Eq.~\eqref{eq:t_effd} into

\begin{equation}\label{eq:t_effd_2}
T_{\rm m} = {\Bigg [} \frac{3}{2} \ \left( \frac{\tau}{2} + \frac{1}{\sqrt{3}} +\frac{1}{3\tau} \right)  \frac{k_{\rm B}}{\sigma_{\rm SB}} \frac{\alpha}{\mu} \Sigma_{\rm p} n {\Bigg ]}^{1/3} \ .
\end{equation}

The mean molecular mass of the gas is determined by the degree of ionization, which can be derived from the Saha equations. For the range of disk temperatures we are interested in ($1\,000\,{\rm K}~{\lesssim}~T~{\lesssim}~6\,000$\,K), the Saha equations predict $1.3~\leq~\mu~\leq~2.4$ \citep[see Eq. 19 and Fig.~2 of][]{2013ApJ...778...77D} in a disk with a composition similar to the protosolar nebula, that is, with hydrogen and helium mass fractions of $X=0.7$ and $Y=0.28$, respectively. We use $\mu=1.85$ in our nominal disk model.

The mass accretion rates of T\,Tauri stars, the variations of FU\,Orionis outbursts, dwarf nova, and X-ray transients suggest $10^{-3}~\lesssim~\alpha~10^{-1}$ \citep{2007MNRAS.376.1740K}. The lower end of this scale likely represents partially ionized disks  while the upper end of this range can be reached in fully ionized disks \citep{2019NewA...70....7M}. The accretion rates and total masses of protoplanetary disks suggest $10^{-3}<\alpha<10^{-2}$ beyond 10\,AU from solar-type stars. In these regions, the disks are only partly ionized (e.g., the outermost layers). The inner parts of a protoplanetary disk (< 0.1\,AU) however can be highly thermally ionized, and so $\alpha> 10^{-2}$ is fairly reasonable. There is also strong evidence that $\alpha$ must be lower than 1 \citep{2019NewA...70....7M} and the substantial ionization of the disk in the hot-Jupiter regime strongly suggests $\alpha>10^{-3}$. All things combined, we chose to test three reference values of $\alpha\in\{10^{-3}, 10^{-2}, 10^{-1}\}$ representing a range of possible disk viscous efficiencies, irrespective of whether they be dominated by eddies in the turbulent gaseous disk or by its magnetic buoyancy \citep{1973A&A....24..337S}. In our nominal disk model we use $\alpha=10^{-3}$.

We model the disk gas surface density as $\Sigma_{\rm p}~=~\Sigma_{\rm p,0}\,a^{-3/2}$ according to the phenomenological minimum-mass solar nebula model \citep{1981PThPS..70...35H,2004ApJ...604..388I}, with a nominal value of $\Sigma_{\rm p,0}=1\,000\,{\rm g\,cm}^{-2}$ at 1\,AU \citep{1997ApJ...486..372B,2012ApJ...755...74K,2013ApJ...779...59G,2017ApJ...835..230F}.

\begin{figure}
\centering
\includegraphics[angle= 0, width=1.\linewidth]{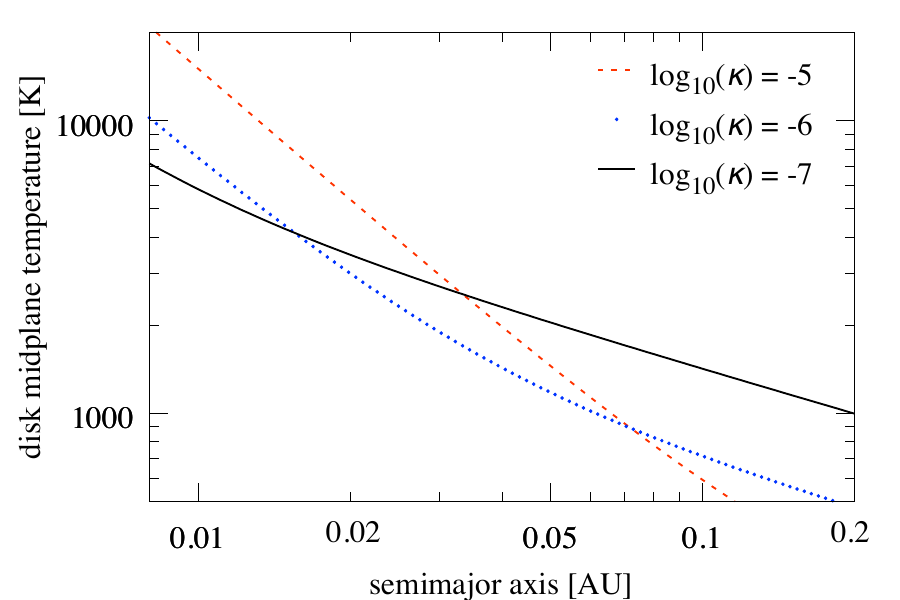}
\caption{Midplane temperature of our disk model, which assumes that viscous heating is the dominant heating term. Examples for three different local disk opacities ($\kappa$, in ${\rm m}^2\,{\rm kg}^{-1}$) are shown. The black line with $\kappa = 10^{-7}{\rm m}^2\,{\rm kg}^{-1}$ is our nominal disk model. With respect to the abscissa, we note that young solar-type stars can have radii of 0.01\,AU (about two solar radii) or more.}
\label{fig:temperature}
\end{figure}

Figure~\ref{fig:temperature} shows $T_{\rm m}(a)$ as per Eq.~\eqref{eq:t_effd_2} for our nominal disk model. Three relations are shown for different disk opacities, $\kappa\in\{10^{-5}, 10^{-6}, 10^{-7}\}\,{\rm m}^2\,{\rm kg}^{-1}$. In the following, we use $\kappa~=~10^{-7}\,{\rm m}^2\,{\rm kg}^{-1}$ because (i) this curve reproduces the 2\,000\,K at 0.05\,AU predicted by \citet{1996Natur.380..606L}; and (ii) it is in good agreement with the midplane temperatures predicted in a numerical 2D rotationally symmetric model by \citet{1997ApJ...486..372B} of a viscous disk for $\alpha\,\approx\,0.001$ and a stellar accretion rate of $10^{-7}\,M_\odot\,{\rm yr}^{-1}$, where $M_\odot$ is the solar mass. We need to keep in mind however that in a more realistic scenario the dust opacity could be modeled as a function of the temperature itself, $\kappa=\kappa(T)$ \citep{1996A&A...311..291H}.

%**********************************************
%Fig. 2
%**********************************************

\begin{figure}
\centering
\includegraphics[angle= 0, width=1.\linewidth]{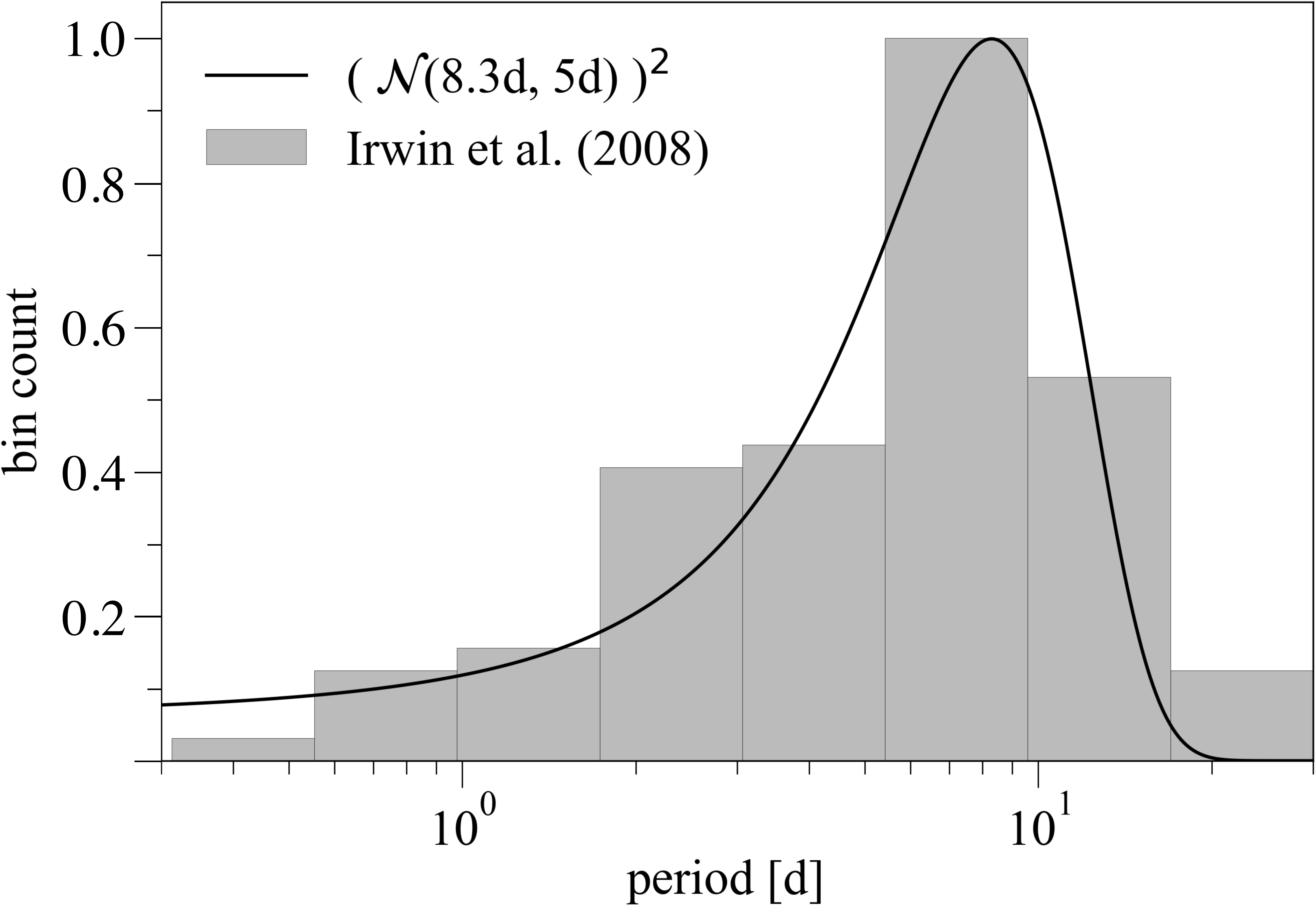}
\caption{Spin periods of stars with masses $0.75 \leq M_\star/M_\odot\leq1.25$ in the 5\,Myr young open cluster NGC\,2362. The histogram shows the data of \citet{2008MNRAS.384..675I}, the solid line is our analytical fit to the data using a squared normal distribution with $\mu=8.3$\,d and $\sigma=5$\,d.}
\label{fig:Prot}
\end{figure}

The stellar rotation determines the corotation radius and can in some scenarios influence the location of the magnetic cavity. In Fig.~\ref{fig:Prot} we show the spin period distribution of stars with masses between $0.75$ and $1.25\,M_\odot$ observed in the 5\,Myr young open cluster NGC\,2362 by \citet{2008MNRAS.384..675I}. The gray bars show the original data and the solid black line shows our fit of a squared normal distribution $(\mathcal{N}(\mu,\sigma))^2$ to the data with $\mu=8.3$\,d as the mean and $\sigma=5$\,d as the standard deviation of $\mathcal{N}$. This rotational period distribution defines a probability density function that we use to randomize the stellar spin period.

Finally, we add the effects of a magnetic inner cavity in the disk, which is reflected in our model as a truncation of the disk torque at the inner 2:1 orbital mean motion resonance with the radial position of the magnetic cavity. The orbital radius at which the disk torque is truncated is located at $0.63\,r_{\rm mc}$.\footnote{The factor of 0.63 is derived in Eq.~\eqref{eq:r_co} in a different context, in which we determine the transition between the dynamical tide and the equilibrium tide at the 2:1 mean motion resonance with the stellar corotation radius $r_{\rm co}= ( \ G(M_\star+M_{\rm p})/\Omega_\star^2 \ )^{1/3}$.} We explore two scenarios for the location of the magnetic cavity. In the first scenario, the magnetic cavity is drawn from a normal distribution parameterized as $r_{\rm mc}=0.05\,(\pm0.015)$\,AU. This scenario is similar to the one proposed by \citet{2002ApJ...574L..87K}, in which the gas disk is truncated at a temperature of 1500\,K by the onset of a magnetorotational instability. \citet{2002ApJ...574L..87K} estimate that this temperature is reached at a distance of about 0.067\,AU. Similarly, \citet{2006ApJ...645L..73R} argued that the critical distance for a cavity to form is the Alfv{\'e}n radius ($r_{\rm A}$), which is at about 0.05\,AU for solar-mass T\,Tauri stars. In a second scenario, we consider that the magnetic cavity is at the same radial distance as the corotation radius.

%**********************************************
%Fig. 3
%**********************************************

\subsection{Torques to drive planet migration}

\subsubsection{The tidal torque}
\label{sec:tidaltorque}

The tidal torque of the star on a nearby planet in its equatorial plane can be estimated as \citep{Efroimsky2013}

\begin{equation}
\Gamma_{\rm t} (a) = \frac{3}{2} G M_{\rm p}^2 k_{2,\star} \frac{R_\star^5}{a^6} \sin(2\epsilon_{\rm g}) \ \ ,
\end{equation}

\noindent
where $k_{2\star}$ is the second-degree tidal Love number of the star and $\epsilon_{\rm g}$ is the instantaneous angular separation between the line connecting the stellar center with the planet and the line connecting the stellar center with the center of the stellar tidal bulges. If we assume that $\epsilon_{\rm g}$ is frequency independent, that the orbital eccentricity is small, and that the planetary orbit is aligned with the stellar equatorial plane, then the tidal torque follows from the quadrupolar modes of the tidal potential and we can introduce a stellar dissipation factor ${Q_\star}$ as per $\sin(2\epsilon_{\rm g})~{\approx}~1/Q_\star$ \citep{murray99}. We can then derive a frequency-averaged, dimensionless quantity for the stellar tidal dissipation as per $k_{2,\star}/\bar{Q}_\star=\langle\mathcal{D}\rangle_\omega$ \citep{2016CeMDA.126..275B}. Thus,

\begin{equation}
\Gamma_{\rm t} (a) = \frac{3}{2} G M_{\rm p}^2 \frac{R_\star^5}{a^6} \langle\mathcal{D}\rangle_\omega \ \ .
\end{equation}

We use the estimates for the dissipation by the stellar dynamical tide \citep{2016CeMDA.126..275B,2016A&A...587A.105A,2017A&A...604A.113B} to calculate $\Gamma_{\rm t} (a)$ on a close-in planet. These stellar models consider frequency-averaged tidal dissipation in the stellar convective envelope \citep{2013MNRAS.429..613O}, which is dominated by the dynamical tide for planets with $|n|<2|\Omega_\star|$ and  by the stellar equilibrium tide otherwise. Variations of the stellar tidal dissipation over short frequency intervals are thus mitigated into the averaged or effective tidal dissipation factor $\bar{Q}_\star$ (for details see Sect.~\ref{sec:numerical}). In our nominal parameterization of a star--planet--disk system, the stellar frequency-averaged tidal dissipation is $\langle\mathcal{D}\rangle_\omega=10^{-3.25}$, a typical value for a solar-type star during the first $\approx\,10$\,Myr of its lifetime \citep{2017A&A...604A.113B}, which corresponds to a frequency-averaged tidal dissipation factor of about $10^{3.4}$. We consider a nominal stellar corotation radius at 0.02\,AU (corresponding to a stellar rotation period of 1\,d), a stellar radius of $2\,R_\odot$, and a Jupiter-mass planet. Beyond this nominal parameterization of the system, which is primarily meant to have an illustrative purpose, we also explore other corotation radii corresponding to the empirically determined rotation period distribution of young stars as per \citet{2008MNRAS.384..675I}.

The algebraic sign of $\Gamma_{\rm t}$ is positive beyond the stellar corotation radius ($a>r_{\rm co}$), where the star transfers its rotational angular momentum to the orbital angular momentum of the planet. In turn, planets closer to the star than the corotation radius transfer angular orbital momentum to spin up the star and, hence, $\Gamma_{\rm t}(a<r_{\rm co})<0$.

\subsubsection{The disk torque}

With $L=M_{\rm p}\sqrt{Ga(M_{\rm p}+M_\star)}$ being the orbital angular momentum of the  planet and without the effects of mass accretion onto the planet (${\rm d}M_{\rm p}/{\rm d}t~{\equiv}~\dot{M}_{\rm p}=0$; $t$ being time), the disk torque in the type-II migration regime is given as

\begin{equation}\label{eq:Gamma_II}
\Gamma_{\rm II} = \frac{{\rm d}L}{{\rm d}t} = L \frac{\dot{a}}{2a}
,\end{equation}

\noindent
where \citep{2004ApJ...604..388I,2005A&A...434..343A}

\begin{equation}\label{eq:dot_a}
\dot{a} = \frac{-3\nu}{2a} \ \min\left( 1 , 2\Sigma_{\rm p} \frac{a^2}{M_{\rm p}} \right) .
\end{equation}

\noindent
We note that in Eq.~\eqref{eq:dot_a}, the disk viscosity depends on the distance to the star, in particular through its coupling with the sound velocity and the midplane temperature in our disk model (Sect.~\ref{sec:disk}).

\subsection{Numerical simulations of tidally driven orbital evolution}
\label{sec:numerical}

\subsubsection{Stellar evolution models}

We also consider the long-term orbital evolution during the tidally dominated period, that is, once the proto-planetary gaseous disk has gone. We use precomputed stellar evolution tracks \citep{2012A&A...543A.108L,2017A&A...604A.113B} that were generated with the STAREVOL code \citep{2016A&A...587A.105A}\footnote{\href{https://obswww.unige.ch/Recherche/evol/starevol/Bolmontetal17.php }{https://obswww.unige.ch/Recherche/evol/starevol/Bolmontetal17.php}\\By courtesy of Florian Gallet (personal communication, May 2018).}. These models consider a 1D rotating star with the radiative core rotating at a different speed than the convective envelope, and they include centrifugal accelerations as well as the resulting chemical stratification. The initial spin period was set to 1.4\,d and the incremental stellar angular momentum loss during each numerical integration time step was calculated according to a differential equation of the torque exerted by magnetic braking with the stellar wind \citep{1997A&A...326.1023B}. The effect of a nearby planet is not taken into account in these models, but according to \citet{2017A&A...604A.113B} the variation due to a hot Jupiter would be limited to about 1\,day in the stellar rotation period after 5\,Gyr. For the critical phase of the tidally driven hot-Jupiter pile-up, which happens on a timescale of 10\,Myr, the tidal effects on the stellar spin can thus safely be neglected. The situation is somewhat different for stars that merge with a Jupiter-mass planet after its tidally driven infall \citep{2012A&A...544A.124B}, but we ignore this effect of star--planet mergers and rather focus on the hot-Jupiter pile-up.

The frequency-averaged tidal dissipation $\langle\mathcal{D}\rangle_\omega$ is calculated during the stellar evolution assuming a simplified two-layer model of the star, that is, a radiative core and a convective envelope. The analytical description developed by \citet{2013MNRAS.429..613O} takes into account the dominating tidal frequencies as a function of the (evolving) stellar properties and is computationally very efficient \citep{Mathis2015}. With $\langle\mathcal{D}\rangle_\omega~=~\int_{-\infty}^{\infty}{\rm d}\omega\,{\rm Im}[k_2^2(\omega)]/\omega$, this procedure is equivalent to calculating the imaginary part of the second-degree tidal Love number of the star along the stellar evolution track.

We use three precomputed stellar evolution tracks with stellar metallicities of ${\rm [Fe/H]} = \log_{10}(Z_\star/Z_\odot)~\in~\{-0.53,0,+0.28\}$, where the solar metal content is given as $Z_\odot\,=\,0.0134$ while $Z_\star\,{\in}\,\{0.004,0.134,0.0255\}$. For comparison, the mean metallicity of stars in the solar neighborhood is ${\rm [Fe/H]} = -0.14\pm0.19$ \citep{2004A&A...418..989N} whereas the stars in the {\it Kepler} field have a near-solar mean metallicity of ${\rm [Fe/H]} = -0.04$ \citep{2017ApJ...838...25G}.

In the stellar evolution models, the star contracts and spins up for approximately the first 100\,Myr until it reaches a minimum rotation of 0.25\,d. In this early phase, the stellar corotation radius moves inward \citep{2012A&A...544A.124B}. Most important for our purpose, the quality factor is calculated consistently from the stellar interior evolution, that is, from the extent of its radiative core and of its convective envelope. This is an important improvement to earlier attempts, which used fixed nominal $Q_\star$ values for particular stages of stellar evolution \citep[see, e.g.,][]{1998ApJ...500..428T}.

\subsubsection{Orbital evolution due to tides}

For our numerical simulations of the tidally driven orbital evolution, we use the precomputed, time-dependent tidal dissipation functions from the stellar evolution tracks and compute the incremental tidal evolution of the planet's orbital semimajor axis ($a$) and orbital eccentricity ($e$) with an adaptive time step ${\rm d}t=10^{-4}t$ as \citep{1998ApJ...499..853E}

%**********************************************
%Fig. 4
\begin{figure*}
\centering
\includegraphics[angle= 0, width=0.492\linewidth]{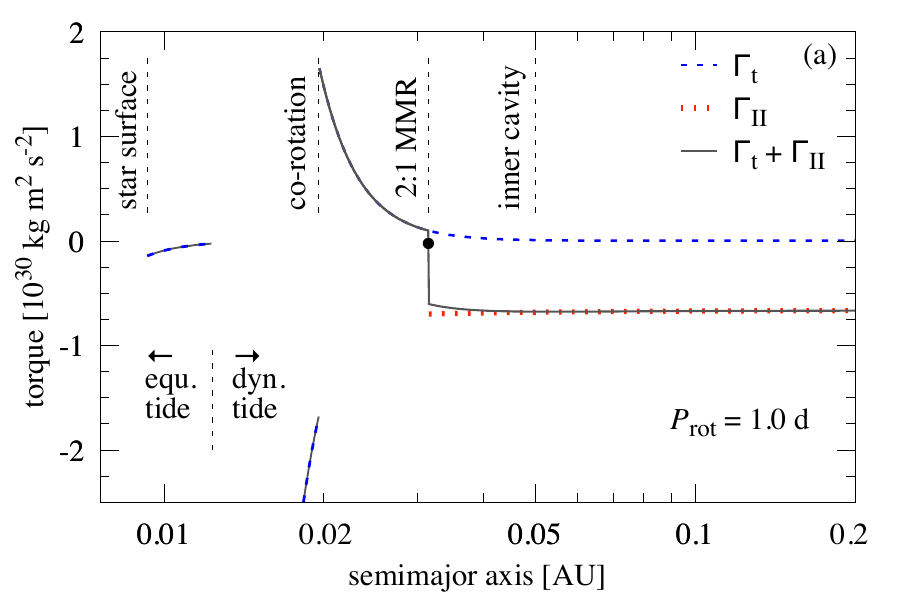}
\hspace{0.05cm}
\includegraphics[angle= 0, width=0.492\linewidth]{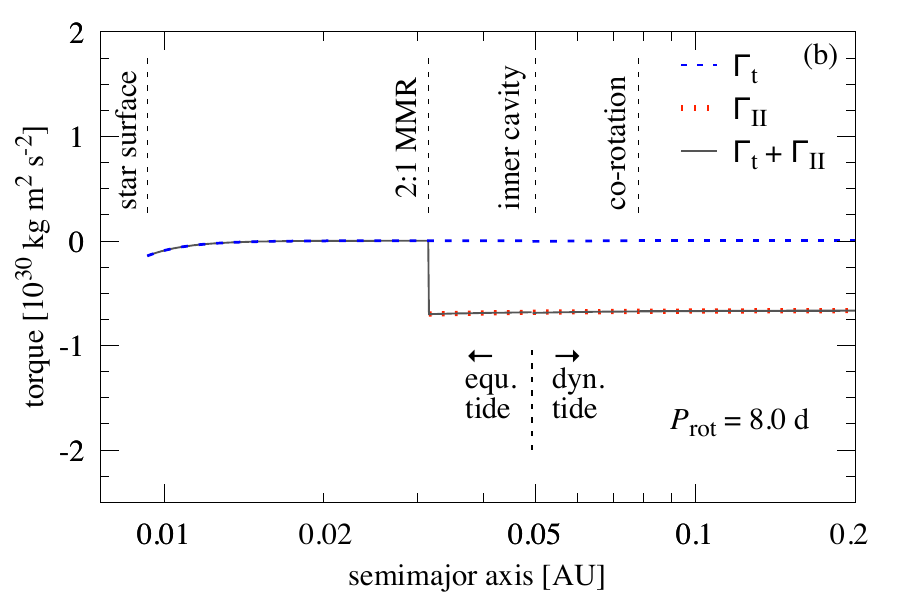}
\caption{Torques exerted on a Jupiter-mass planet embedded in a proto-planetary disk around a young, solar-type star. \textbf{(a)} Nominal parameterization of the star--planet--disk model, with $R_\star=2\,R_\odot$, $P_{\rm rot}=1$\,d, $\langle\mathcal{D}\rangle_\omega=10^{-3.25}$, $\Sigma_{\rm p,0} = 1\,000\,{\rm g\,cm}^{-2}$, $\alpha=10^{-3}$, $\mu = 1.85$, $\kappa=10^{-7}\,{\rm m}^2\,{\rm kg}^{-1}$, and $r_{\rm mc}=0.05$\,AU. The blue dashed curve refers to the tidal torque ($\Gamma_{\rm t}$), the red dotted line to the type-II migration torque of the disk ($\Gamma_{\rm II}$), and the thick solid line to the total torque. The location of zero total torque at 0.031\,AU is indicated with a black circle. The factors $\Gamma_{\rm t}$, and therefore also $\Gamma_{\rm t}+\Gamma_{\rm II}$, swap signs at the stellar corotation radius, interior to which any planet would be rapidly pulled into the star. \textbf{(b)} Similar to \textbf{(a)} but with $P_{\rm rot}=8$\,d.}
\label{fig:torque}
\end{figure*}
%**********************************************

\begin{align} \label{eq:da}
{\rm d}a &= -{\rm d}t  \ a \ \sum_{i=\star,{\rm p}} \frac{1}{T_i} {\Bigg [} \frac{f_1(e)}{\beta^{15}} - \frac{f_2(e)}{\beta^{12}} \frac{\Omega_i}{n}{\Bigg ]} \\
{\rm d}e &= -{\rm d}t  \ e \ \sum_{i=\star,{\rm p}} \frac{9}{2T_i} {\Bigg [} \frac{f_3(e)}{\beta^{13}} - \frac{11 f_4(e)}{18\beta^{10}} \frac{\Omega_i}{n}{\Bigg ]} \label{eq:de} \ ,
\end{align}

\noindent
where \citep{1981A&A....99..126H,2010ApJ...723..285H}

\begin{align}\label{eq:f_e}
\nonumber
\beta(e) & = \sqrt{1-e^2} \ ,\\
\nonumber
f_1(e) & = 1 + \frac{31}{2} e^2 + \frac{255}{8} e^4 + \frac{185}{16} e^6 + \frac{25}{64} e^8 \\
\nonumber
f_2(e) & = 1 + \frac{15}{2} e^2 + \frac{45}{8} e^4 \ \ + \frac{5}{16} e^6 \\
\nonumber
f_3(e) & = 1 + \frac{15}{4} e^2 + \frac{15}{8} e^4 \ \ + \frac{5}{64} e^6 \\
f_4(e) & = 1 + \frac{3}{2} e^2 \ \ + \frac{1}{8} e^4,
\end{align}

\noindent
and

\begin{align}
T_i &= \frac{1}{9} \frac{M_i}{M_j(M_\star+M_{\rm p})} \frac{a^8}{R_i^{10}} \frac{1}{\sigma_i} \ \ \ i \in \{\star,{\rm p}\} \ni j, i \neq j, \label{eq:dyntid}
\end{align}

\noindent
with

\begin{align}
\sigma_\star &= \frac{1}{3} \frac{G}{R_\star^5} |n-\Omega_\star|^{-1} \langle\mathcal{D}\rangle_\omega \ \ , n<|2\Omega_\star| \ {\rm (dynamical \, tide)} \label{eq:dyn} ,\\
\sigma_\star &= \sigma_{0,\star} \ \bar{\sigma}_\star \hspace{2.cm} , n \geq |2\Omega_\star| \ {\rm (equilibrium \, tide)}  \label{eq:equ} ,\\
\sigma_{\rm p} &= \sigma_{0,{\rm p}} \ \bar{\sigma}_{\rm p} ,\\
\sigma_{0,\star} &= \sqrt{G/(M_\star R_\star^7)} \ , \ \bar{\sigma}_\star = 3 \times 10^{-7} \label{eq:sigma0s} ,\\
\sigma_{0,{\rm p}} &= \sqrt{G/(M_{\rm p} R_{\rm p}^7)} \ , \ \bar{\sigma}_{\rm p} = 1 \times 10^{-7,} \label{eq:sigma0p}
\end{align}

\noindent
and with $G$ as the gravitational constant.

Equations~\eqref{eq:da} and \eqref{eq:de} are valid for arbitrary eccentricities \citep{2010A&A...516A..64L} and we tested systems with small initial eccentricities that did not result in qualitatively different behavior of the orbital evolution. Hence, we focus this report on $e=0$ and assume that the stellar and planetary spin axes are aligned.

We read $\langle\mathcal{D}\rangle_\omega$ from the precomputed stellar evolution models based on the frequency-averaged analytical expressions \citep{2013MNRAS.429..613O,Mathis2015}. The frequency-averaged tidal dissipation constant of the star is calculated as $\bar{Q}_\star=3/(2\langle\mathcal{D}\rangle_\omega)$. Equations~\eqref{eq:dyn} and \eqref{eq:equ} ensure that the planet excites tidal inertial waves in the stellar convective layer as long as $n<|2\Omega_\star|$, whereas tidal friction is more adequately modeled by the equilibrium tide for more close-in planets with $n \geq |2\Omega_\star|$ \citep{2016CeMDA.126..275B}. Converting orbital and spin frequencies into orbital radii, Kepler's third law of planetary motion predicts the transition radius ($r_{{\rm e}\leftrightarrow{\rm d}}$) between the equilibrium tide and the dynamical tide regime as

\begin{equation}\label{eq:r_co}
r_{{\rm e}\leftrightarrow{\rm d}} = {\Bigg (} \frac{ G(M_{\rm p}~+~M_\star) }{ (2\Omega_\star)^2 }{\Bigg )}^{1/3} = {\Bigg (}\frac{1}{2}{\Bigg )}^{2/3}\,r_{\rm co} \approx 0.63\,r_{\rm co} \ .
\end{equation}

\noindent
The calibrated tidal dissipation constants in equations~\eqref{eq:sigma0s} and \eqref{eq:sigma0p} are taken from \citet{2010ApJ...723..285H,2012ApJ...757....6H}.

In order to compare this model to a pure equilibrium tide model with fixed $Q_\star$ and constant stellar rotation, we set up another suite of simulations, which assumes a sun-like stellar radius, mass, and rotation, and estimates $\sigma_\star$ in Eq.~\eqref{eq:dyntid} as

\begin{equation}
\sigma_\star = \frac{G \, k_{2,\star}^2}{|n-\Omega_\star| Q_\star R_\star^5} \ ,
\end{equation}

\noindent
an approximation that is only valid in the limit of small eccentricities \citep{2016CeMDA.126..275B}.

%**********************************************
%Fig. 5
\begin{figure*}
\centering
\includegraphics[angle= 0, width=0.50\linewidth]{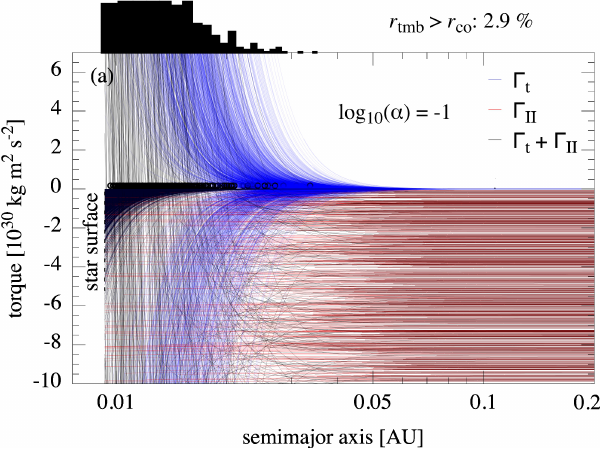}
\hspace{-0.05cm}
\includegraphics[angle= 0, height=0.37\linewidth]{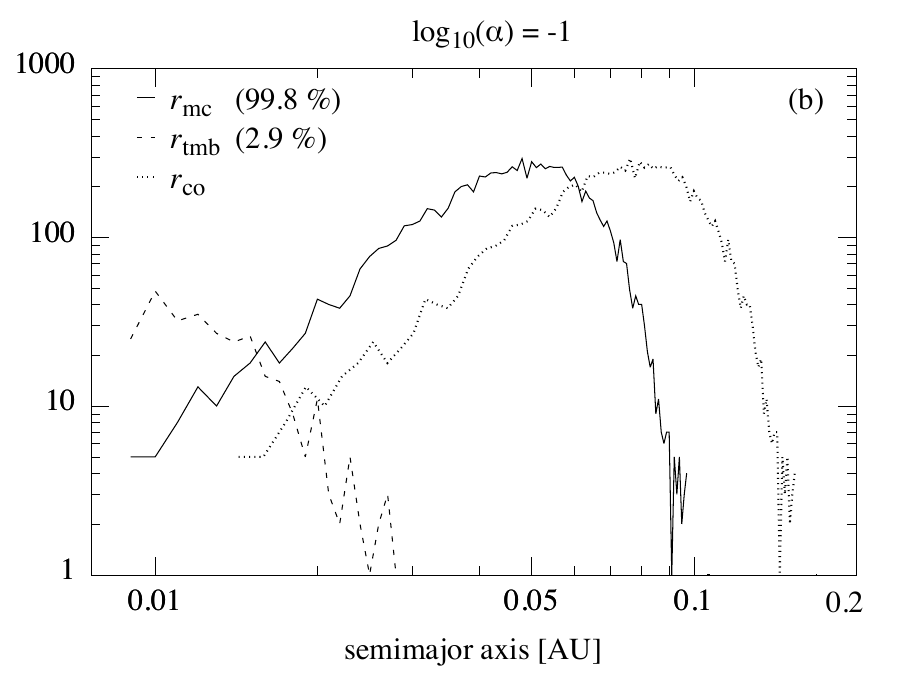}
\caption{\textbf{(a)} Monte Carlo simulations of 10,000 different disks (with $\alpha=10^{-1}$) and stars with stellar rotations periods drawn from a probability density distribution derived from \citet{2008MNRAS.384..675I} in Fig.~\ref{fig:Prot}. A tidal migration barrier existed in 3\,\% of all simulations and their locations are shown in the black histogram at the top of the panel. The location of the magnetic cavity $r_{\rm mc}$ was different in each simulation, depending on the stellar rotation period. \textbf{(b)} Histograms of the radial distance of the magnetic cavity ($r_{\rm mc}$), which existed in 90\,\% of all simulations, and of the tidal migration barrier ($r_{\rm tmb}$). The corotation radius distribution is plotted at 1/0.63 times $r_{\rm mc}$.}
\label{fig:Prot_rand_alpha-1_rmc_RAND}
\end{figure*}
%**********************************************

%**********************************************
%Fig. 6
\begin{figure*}[h!]
\centering
\includegraphics[angle= 0, width=0.50\linewidth]{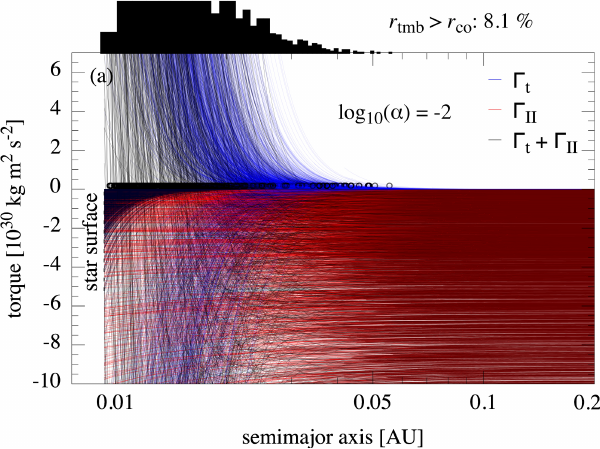}
\hspace{-0.05cm}
\includegraphics[angle= 0, height=0.37\linewidth]{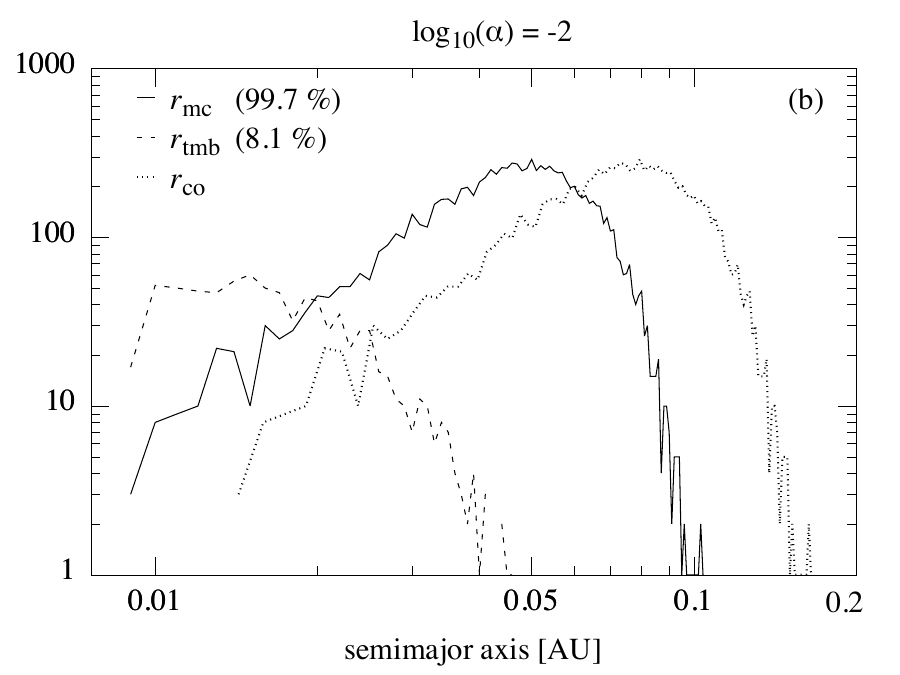}
\caption{Same as Fig.~\ref{fig:Prot_rand_alpha-1_rmc_RAND} but for $\alpha=10^{-2}$. The tidal migration barrier exists in 8\,\% and the magnetic cavity in 90\,\% of the simulations.}
\label{fig:Prot_rand_alpha-2_rmc_RAND}
\end{figure*}
%**********************************************

%**********************************************
%Fig. 7
\begin{figure*}[h!]
\centering
\includegraphics[angle= 0, width=0.50\linewidth]{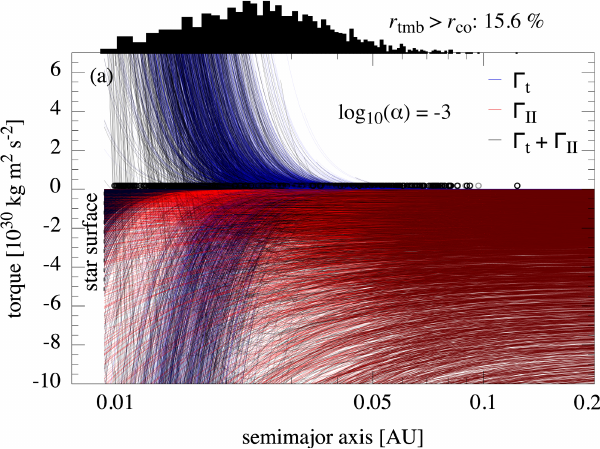}
\hspace{-0.05cm}
\includegraphics[angle= 0, height=0.37\linewidth]{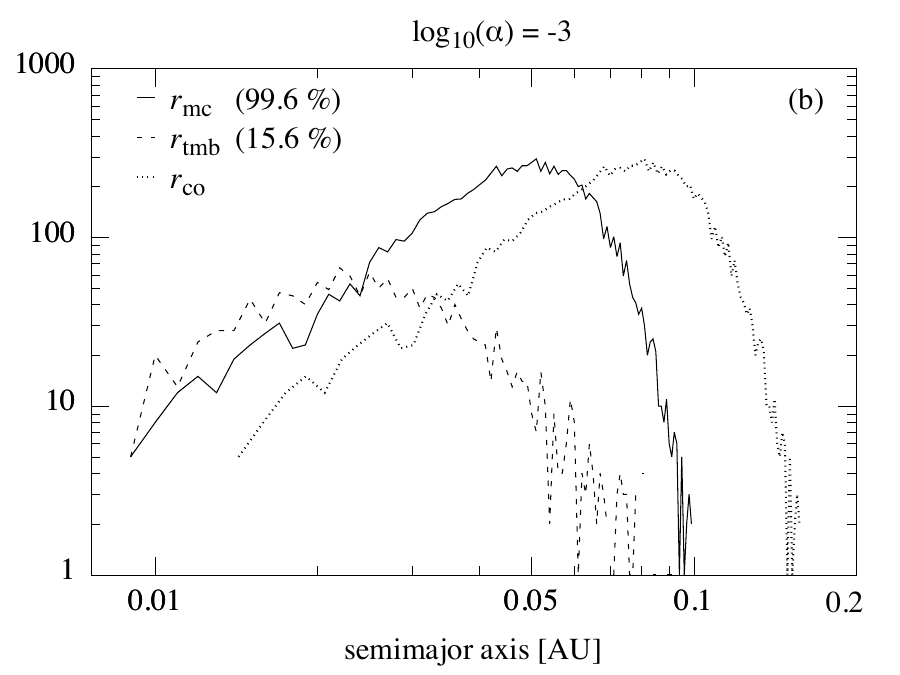}
\caption{Same as Fig.~\ref{fig:Prot_rand_alpha-1_rmc_RAND} but for $\alpha=10^{-3}$. The tidal migration barrier exists in 15\,\% and the magnetic cavity in 90\,\% of the simulations.}
\label{fig:Prot_rand_alpha-3_rmc_RAND}
\end{figure*}
%**********************************************

%**********************************************
%Fig. 8
\begin{figure}[h!]
\centering
\includegraphics[angle= 0, width=1.0\linewidth]{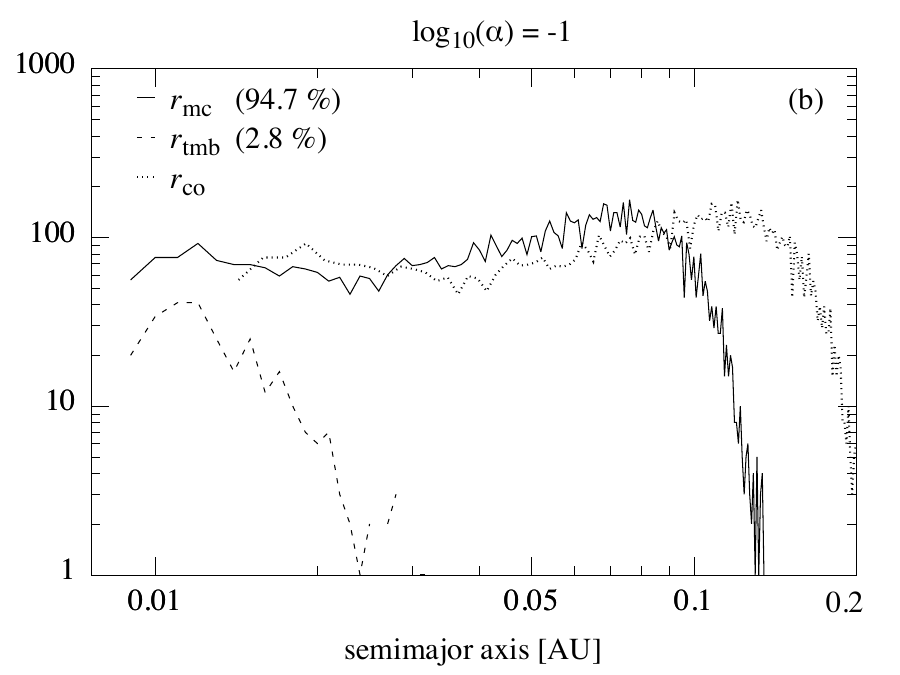}\\
\includegraphics[angle= 0, width=1.0\linewidth]{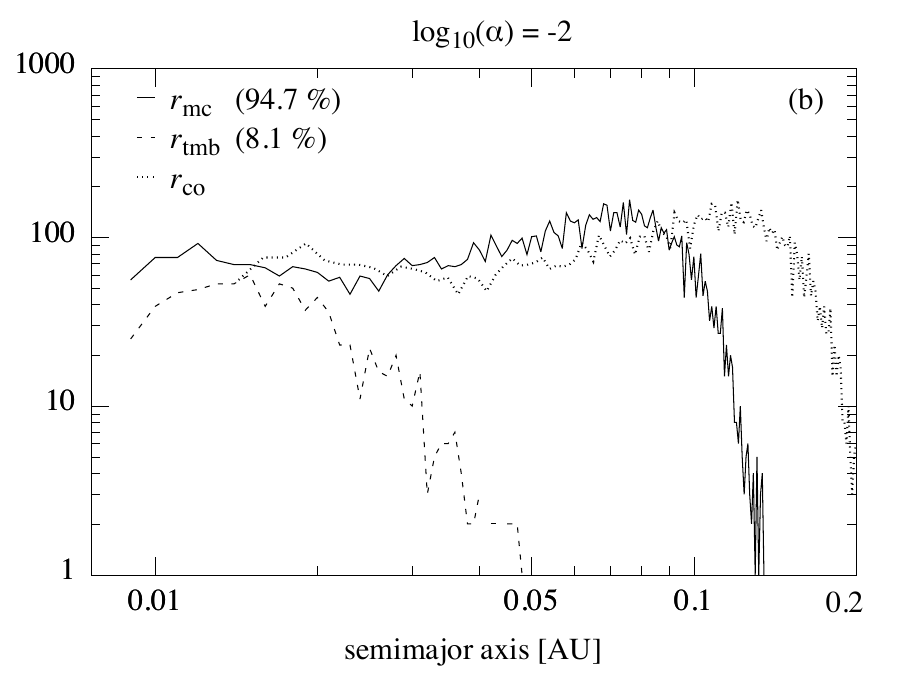}\\
\includegraphics[angle= 0, width=1.0\linewidth]{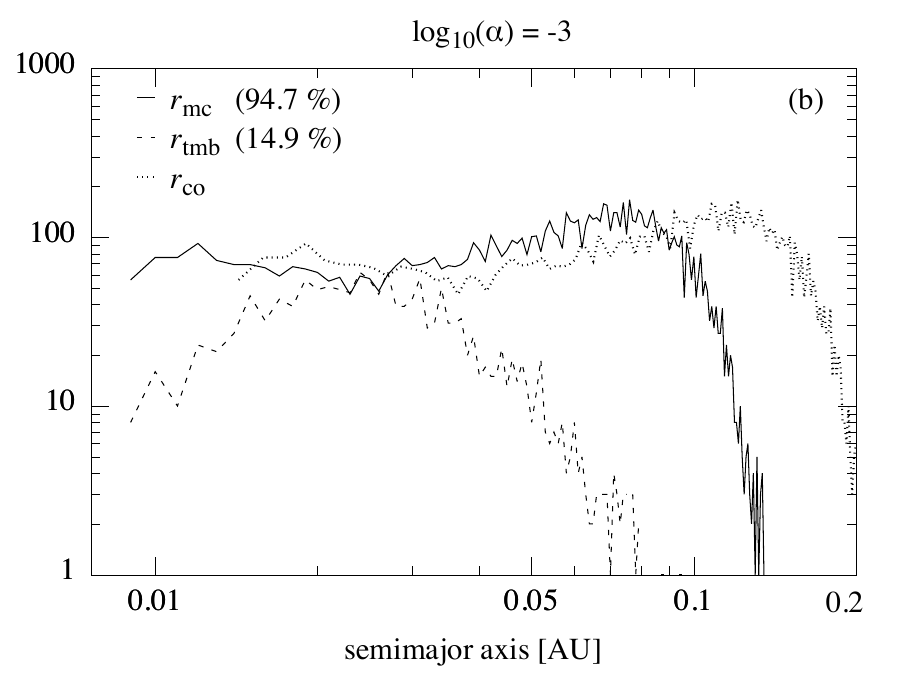}
\caption{Same as Figs.~\ref{fig:Prot_rand_alpha-1_rmc_RAND}b, \ref{fig:Prot_rand_alpha-2_rmc_RAND}b, and \ref{fig:Prot_rand_alpha-3_rmc_RAND}c except that the magnetic cavity has been modeled to coincide with the corotation radius.}
\label{fig:Prot_rand_rmc_rco}
\end{figure}
%**********************************************

\section{Results}
\label{sec:results}

\subsection{Tidal migration barrier at zero torque}

%**********************************************
%Fig. 9
\begin{figure*}
\centering
\includegraphics[angle= 0, width=0.5\linewidth]{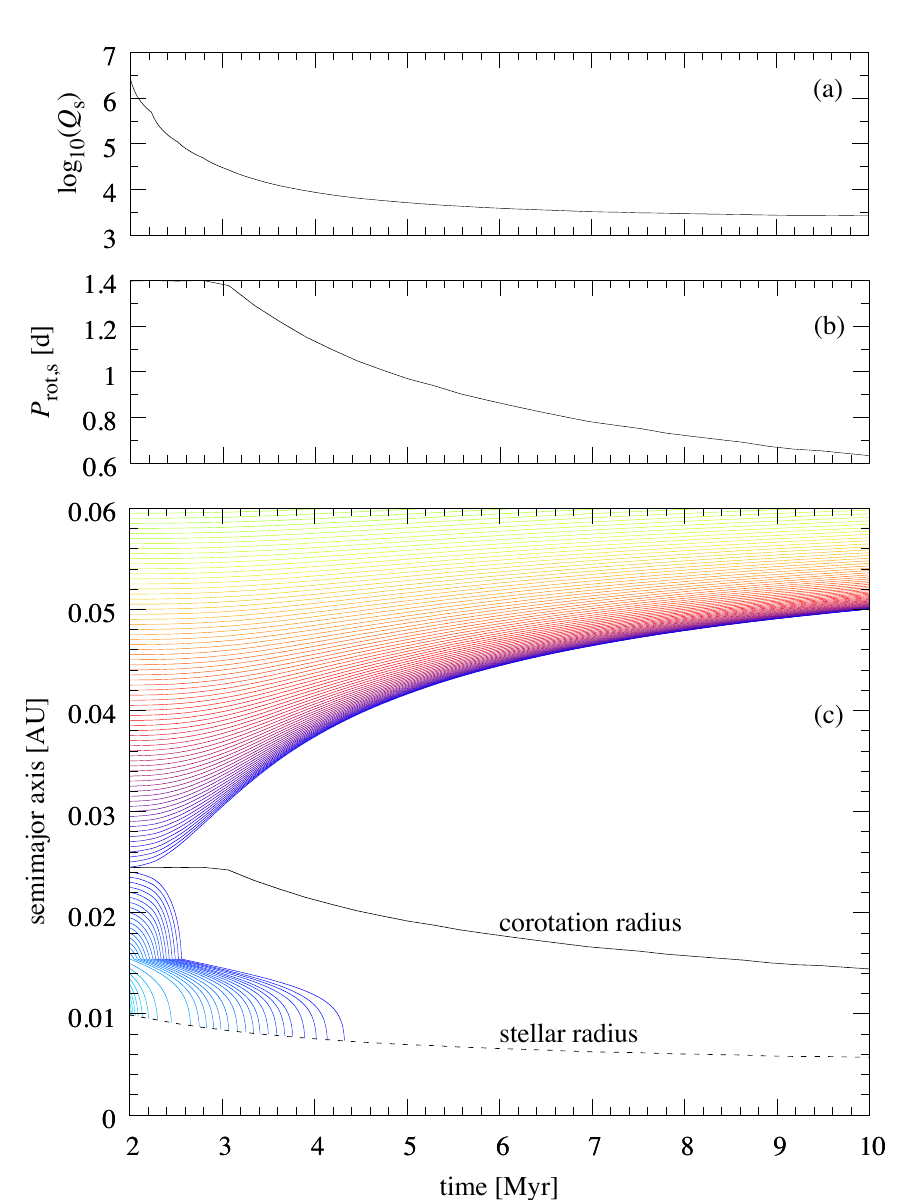}
\includegraphics[angle= 0, width=0.493\linewidth]{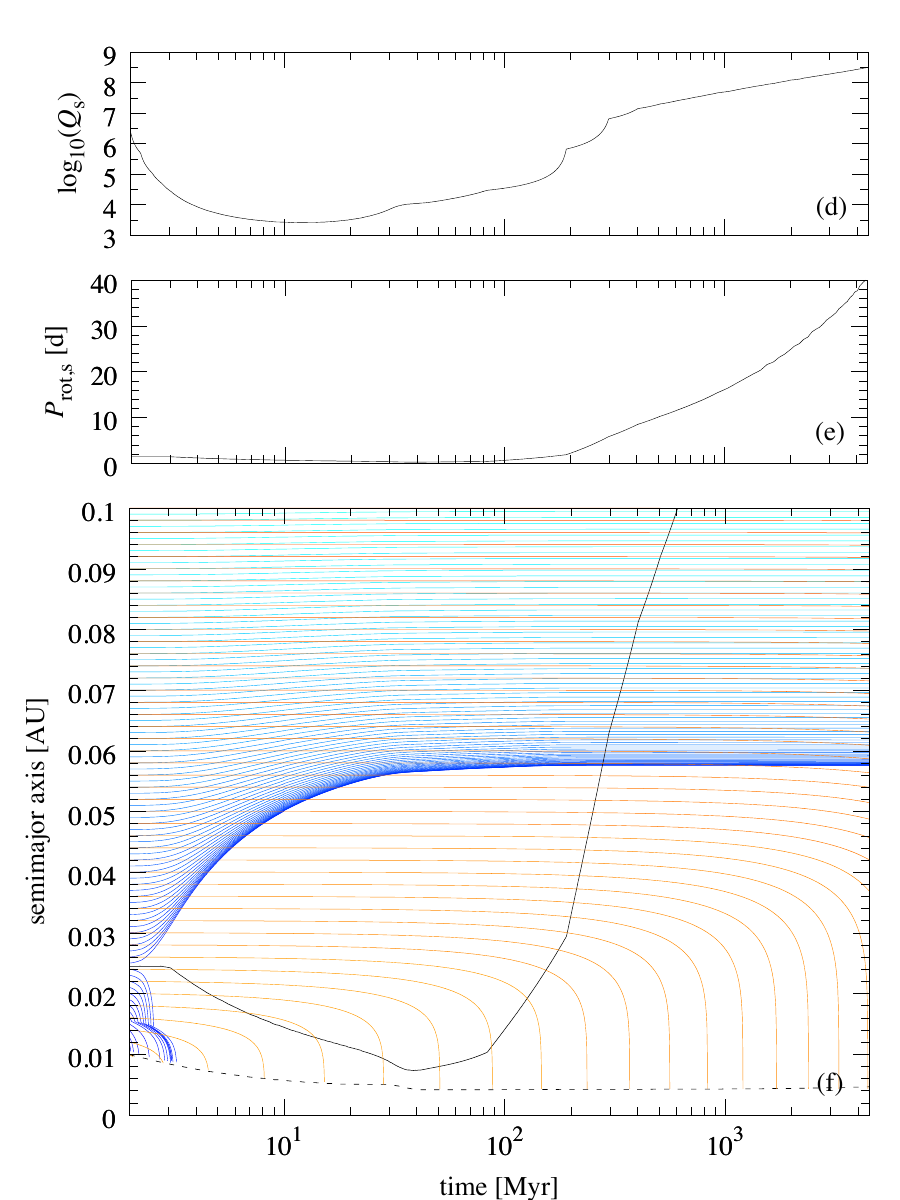}
\caption{Evolution of the spin and orbital properties of a Sun-like star (initial rotation period 1.4\,d, metallicity $Z=0.0134$) as per \citet{2016A&A...587A.105A} and \citet{2016CeMDA.126..275B} and orbital evolution of a hot-Jupiter population. \textbf{(a)} Frequency-averaged tidal dissipation factor and \textbf{(b)} rotation period of the star during the first 10\,Myr of stellar evolution. \textbf{(c)} Tidally driven orbital evolution of a single planet on a grid of 100 equally spaced initial orbits. Orbital decay is calculated via Eq.~\eqref{eq:da} (assuming $e=0$) according to the dynamical tide model with stellar evolution as per \textbf{(a)} and \textbf{(b)}. \textbf{(d)} Stellar dissipation factor and \textbf{(e)} stellar rotation period over the first 1\,Gyr of stellar evolution. \textbf{(f)} Comparison of the planetary orbital evolution in the dynamical tide model (blue lines) and in the equilibrium tide model (orange lines, $Q_\star~=~10^5$).}
\label{fig:stelltid_evol}
\end{figure*}
%**********************************************

%**********************************************
%Fig. 10
\begin{figure*}
\centering
\includegraphics[angle= 0, width=0.492\linewidth]{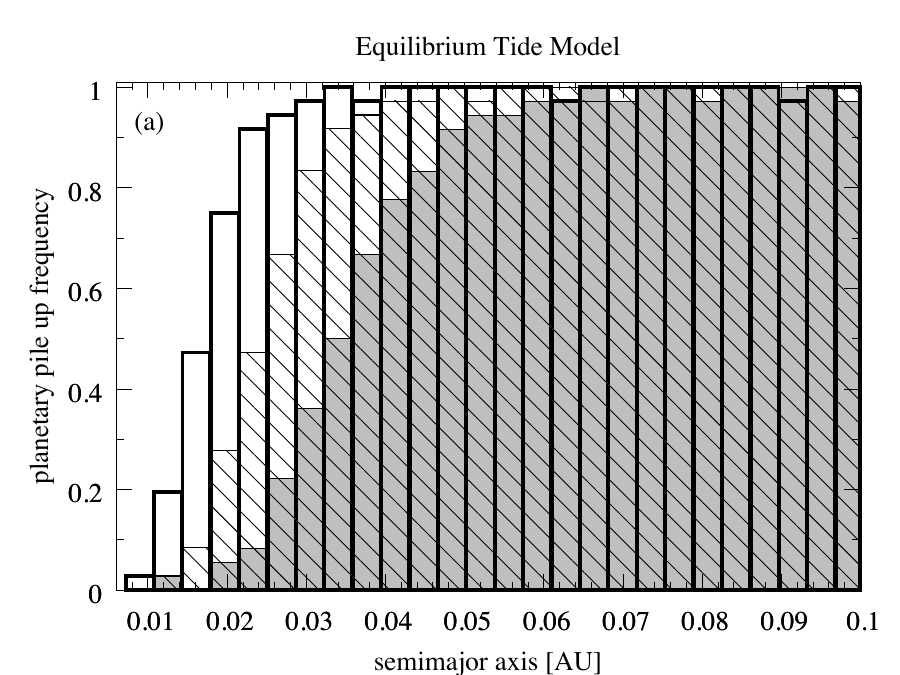}
\hspace{0.02cm}
\includegraphics[angle= 0, width=0.492\linewidth]{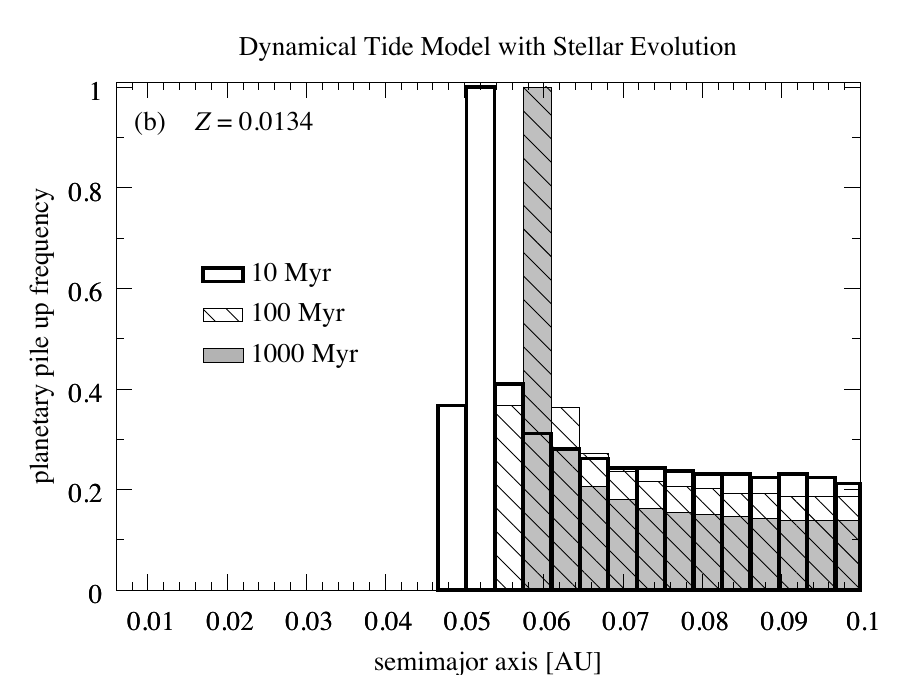}
\includegraphics[angle= 0, width=0.492\linewidth]{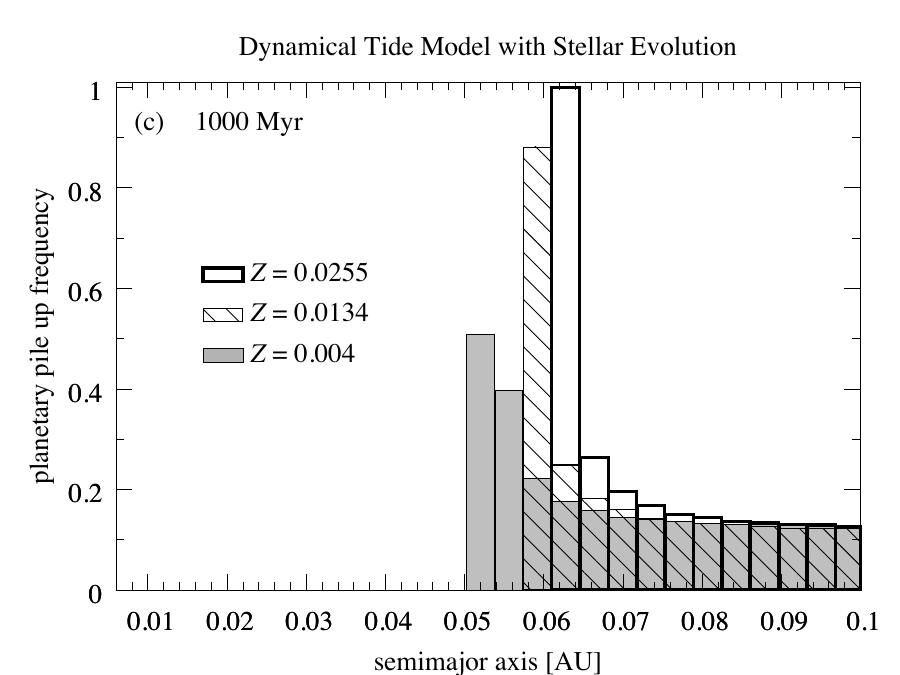}
\caption{Normalized histograms of 911 orbital integrations of a Jupiter-mass planet around a Sun-like star as per Fig.~\ref{fig:stelltid_evol}(f). \textbf{(a)} assumes the equilibrium tide model and a fixed $Q_\star=10^5$, \textbf{(b)} is based on the dynamical tide model with stellar evolution of a sun-like star (metallicity $Z=0.0134$). In \textbf{(a)} and \textbf{(b)}, each histogram is normalized to a maximum of 1 and different shadings refer to different integration times of our numerical code; see legend in \textbf{(b)}. In \textbf{(c)}, different histogram shadings refer to simulated hot-Jupiter populations around stars in the dynamical tide model and stellar evolution with sub-solar ($Z=0.004$), solar ($Z=0.0134$), and super-solar ($Z=0.0255$) metallicities after 1000\,Myr. These histograms are scaled to agree at 0.1\,AU, beyond which tides become insignificant.}
\label{fig:stelltid_hist}
\end{figure*}
%**********************************************

Figure~\ref{fig:torque} displays the total torque $\Gamma_{\rm tot}~{\equiv}~\Gamma_{\rm t} (a)+\Gamma_{\rm II} (a)$ (black solid line) acting on a close-in Jupiter-mass planet in our nominal disk scenario and for a Sun-like star with its corotation radius at 0.02\,AU. The purpose of Fig.~\ref{fig:torque} is to illustrate the general behavior of the torques for different possible radial arrangements of the inner magnetic cavity of the disk (here fixed at 0.05\,AU) and the stellar corotation radius. The 2:1 MMR with the cavity determines the truncation of the disk torque acting on the planet. The corotation radius determines the transition from the dynamical tide to the equilibrium tide regime. All things combined determine the existence and, as the case may be, the radial position of the tidal migration barrier. In Fig.~\ref{fig:torque}(a), $P_{\rm rot}=1$\,d and the corotation radius is interior to the magnetic cavity. In Fig.~\ref{fig:torque}(b), $P_{\rm rot}=8$\,d and the inner cavity happens to be at 0.63 times the corotation radius, that is, the inner cavity is in a 2:1 MMR with $r_{\rm co}$.

In Fig.~\ref{fig:torque}(a), the negative type-II migration torque (red dotted line) dominates the positive torque from the stellar dynamical tide (blue dashed line) beyond the 2:1 MMR with the inner cavity, so that $\Gamma_{\rm tot}<0$ and the planet migrates inward. Interior to the 2:1 MMR with the cavity, the disk torque is zero and so the total torque (black line) switches its algebraic sign according to the positive (i.e., repulsive) torque of the stellar dynamical tide. As a consequence, the total torque must go through a point of zero torque. This location of zero torque, which we refer to as the tidal migration barrier, is indicated with a black dot and it is the distance at which planet migration stops.

Interior to the tidal migration barrier, where the planet cannot arrive via migration alone, the strong distance dependence of $\Gamma_{\rm t} (a)~{\propto}~a^{-6}$ leads to a very effective tidal repulsion. Subsequently, at the corotation radius, the algebraic sign of the tidal torque (and therefore of the total torque as well) switches sign to negative and the planet falls towards the star relatively fast. At the 2:1 MMR with the corotation radius however, the stellar tides transition from the dynamical tide to the equilibrium tide, which is much weaker than the dynamical tide. As a consequence, the magnitude of the negative tidal torque is reduced as well. This transition is located at 0.02\,AU~$\times~0.63=0.0126$\,AU (see Eq.~\ref{eq:r_co}) in Fig.~\ref{fig:torque}(a). It is very possible that hot Jupiters, which do not encounter a tidal migration barrier beyond $r_{\rm co}$ may survive in this extremely close stellar vicinity on a timescale of 10\,Myr - 100\,Myr if they can also withstand tidal disruption.

In Fig.~\ref{fig:torque}(b), where $P_{\rm rot}=8$\,d, the corotation radius is beyond the inner magnetic cavity of the disk, which is at 0.05\,AU as in Fig.~\ref{fig:torque}(a). The stellar rotation period was intentionally chosen to be 8\,d so that the transition between the dynamical tide and the equilibrium tide regime coincides with the radial position of the inner cavity, although it is entirely unclear at this point if the magnetic cavity can be affected directly by the stellar rotation at all.

Beyond our nominal parameterization of the star--planet--disk system in Fig.~\ref{fig:torque}, we explore a range of possible realizations in Figs.~\ref{fig:Prot_rand_alpha-1_rmc_RAND}-\ref{fig:Prot_rand_alpha-3_rmc_RAND}. We keep the stellar radius and the planetary mass fixed, while we draw $\langle\mathcal{D}\rangle_\omega$ from a log-normal distribution as per $\log_{10}(\langle\mathcal{D}\rangle_\omega)~=~-3.25\,\pm0.5$. Similarly, we vary $\log_{10}(\Sigma_{\rm p,0} / [{\rm g\,cm}^{-2}]) = 3\,\pm1$ and draw the mean molecular weights from a normal distribution as per $\mu~=~1.85~\pm~0.55$. We also randomize $P_{\rm rot}$ as per the probability density distribution shown in Fig.~\ref{fig:Prot} and also randomly draw the location of the magnetic cavity from a normal distribution parameterized as $r_{\rm mc}=0.05\,(\pm0.015)$\,AU (see Sect.~\ref{sec:disk}).

We generate 10,000 Monte Carlo simulations for $\alpha=10^{-1}$ (Fig.~\ref{fig:Prot_rand_alpha-1_rmc_RAND}), $\alpha=10^{-2}$ (Fig.~\ref{fig:Prot_rand_alpha-2_rmc_RAND}), and $\alpha=10^{-3}$ (Fig.~\ref{fig:Prot_rand_alpha-3_rmc_RAND}), respectively. These randomized parameterizations are carried out to derive a plausible distribution of the tidal migration barrier for hot Jupiters around sun-like stars. The resulting rates of a tidal migration barrier outside of the corotation radius are 2.9\,\% for $\alpha=10^{-1}$, 8.1\,\% for $\alpha=10^{-2}$, and 15.6\,\% for $\alpha=10^{-3}$. Panels (a) in Figs.~\ref{fig:Prot_rand_alpha-1_rmc_RAND}-\ref{fig:Prot_rand_alpha-3_rmc_RAND} show the distribution of the corresponding tidal, disk, and total torques in analogy to Fig.~\ref{fig:torque}. Panels (b) in Figs.~\ref{fig:Prot_rand_alpha-1_rmc_RAND}-\ref{fig:Prot_rand_alpha-3_rmc_RAND} illustrate the distributions of the magnetic cavity (solid lines), tidal migration barrier (dashed lines), and corotation radius (dotted line). In comparing Figs.~\ref{fig:Prot_rand_alpha-1_rmc_RAND}-\ref{fig:Prot_rand_alpha-3_rmc_RAND}, we highlight that the histograms of $r_{\rm mc}$ and $r_{\rm co}$ do not depend on the respective choice of $\alpha$, while the distribution of $r_{\rm tmb}$ does depend on $\alpha$. The only way for a magnetic cavity not to exist in our simulations is if it is located within the stellar radius.

As an alternative scenario for the location of the magnetic cavity, we consider that the magnetic cavity coincides with the corotation radius. In Fig.~\ref{fig:Prot_rand_rmc_rco} we plot the histograms of $r_{\rm mc}$, $r_{\rm tmb}$, and $r_{\rm co}$ from the resulting Monte Carlo simulations. We find that the rates for the tidal migration barrier being located beyond the corotation radius are very similar in all three cases of the $\alpha$ viscosity parameter that we studied as summarized in the upper left corners of the panels in Fig.~\ref{fig:Prot_rand_rmc_rco}.

With $f_{\rm occ}=1.2\,\pm0.38\,\%$ as the bias-corrected observed occurrence rate \citep{2012ApJ...753..160W}, $f_{\rm sur}\in\{2.9\,\%, 8.1\,\%, 15.6\,\%\}$ as the survival rates from our simulations, $f_{\rm alt}$ as the rate of all alternative mechanisms to stop a migrating Jupiter from falling into its star, and $f_{\rm for}$ as the hot-Jupiter formation frequency around sun-like stars, we have $f_{\rm occ}=(f_{\rm sur} + f_{\rm alt})\,\cdot\,f_{\rm for}$. This is equivalent to $f_{\rm for}=f_{\rm occ}/(f_{\rm sur}+f_{\rm alt})<f_{\rm occ}/f_{\rm sur}$, for which we can estimate $f_{\rm sur}$ from our simulations.

In the first scenario, where the radial position of the magnetic cavity is chosen from a normal distribution, we have $f_{\rm for}<41.4\,\pm13.1\,\%$ for $\alpha=10^{-1}$, $f_{\rm for}=14.8\,\pm4.7\,\%$ for $\alpha=10^{-2}$, and $f_{\rm for}=7.7\,\pm2.4\,\%$ for $\alpha=10^{-3}$. In the second scenario, where the magnetic cavity coincides with the corotation radius, we find $f_{\rm for}<42.9\,\pm13.6\,\%$ for $\alpha=10^{-1}$, $f_{\rm for}=14.8\,\pm4.7\,\%$ for $\alpha=10^{-2}$, and $f_{\rm for}=8.1\,\pm2.6\,\%$ for $\alpha=10^{-3}$. Hence, in both scenarios the resulting hot-Jupiter survival rates and the formation rates are very similar. In other words, when combined with observations, our simulations suggest that less than about 8\,\% ($\alpha=10^{-3}$) to 43\,\% ($\alpha=10^{-1}$) of sun-like stars initially encounter an inward migrating hot Jupiter. These values are almost irrespective of whether the magnetic cavity is located at about 0.05\,AU and independent of the stellar rotation or of whether or not the magnetic cavity is coupled to the corotation radius. All these results are summarized in Table~\ref{tab:rates}.

\subsection{Hot-Jupiter pile up from tidally driven migration}

Once the protoplanetary disk has been accreted onto the central star after about 10\,Myr into the star's lifetime \citep{2001ApJ...553L.153H}, the disk torque vanishes and the orbit of a hot Jupiter evolves under the effect of stellar tides only, neglecting the possibility of interaction with other planets or nearby stars. Figure~\ref{fig:stelltid_evol} shows the outcome of our numerical simulations for a solar metallicity star, which are based on differential equations for d$a$/d$t$ (assuming $e=0$) as derived from the orbit-averaged torque for a tidally evolving two-body system. Panels (a)-(c) on the left illustrate the first 10\,Myr of evolution, and panels (d)-(f) on the right side show the evolution over 4.5 billion years, that is, over the age of the solar system. Figures~\ref{fig:stelltid_evol}(a) and (d) demonstrate the variation of $\bar{Q}_\star$ over several orders of magnitude. The initial $\bar{Q}_\star\approx10^{3.25}$ implies highly effective dissipation over some 10\,Myr, whereas values of $\bar{Q}_\star>10^8$ after one billion years signify negligible dissipation. The initial spin-up due to contraction and the subsequent spin-down owing to magnetic braking are displayed in panels (b) and (e). The orbital evolution of 100 star--planet two-body systems is shown in panels (c) and (f).

\begin{table*}
\caption{Hot-Jupiter survival rates at the tidal migration barrier and resulting formation rates in our model for various parameterizations of the protoplanetary disk. Formation rates are deduced from the bias-corrected hot-Jupiter occurrence rate of $1.2\,\pm0.38\,\%$ \citep{2012ApJ...753..160W}.}
\def\arraystretch{1.4}
\label{tab:rates}
\centering
\begin{tabular}{c | c c c  | ccc}
\hline\hline
 & \multicolumn{3}{c}{$r_{\rm mc}$ randomized\tablefootmark{a}}  & \multicolumn{3}{c}{$r_{\rm mc}=r_{\rm co}$}\\
$\log(\alpha)$ & $-1$  & $-2$ & $-3$ & $-1$ & $-2$ & $-3$\\
\hline
$f_{\rm sur}$ & 2.9\,\% & 8.1\,\% & 15.6\,\% & 2.8\,\% & 8.1\,\% & 14.9\,\%\\
$f_{\rm for}$ & $41.4\,\pm13.1$\,\% & $14.8\,\pm4.7$\,\% & $7.7\,\pm2.4$\,\% & $42.9\,\pm13.6$\,\% & $14.8\,\pm4.7$\,\% & $8.1\,\pm2.6$\,\%\\
\hline
\end{tabular}
\tablefoot{ \tablefoottext{a}{Our randomizations of the radial position of the magnetic cavity assume a normal distribution with mean value of 0.05\,AU and a standard deviation of 0.015\,AU.}}
\end{table*}

In Fig.~\ref{fig:stelltid_evol}(c), colored lines refer to the orbital evolution of the planet, with color encoding the initial orbital semi-major axis. Three mechanisms are readily visible: (1) the pile-up of hot Jupiters at about 0.05\,AU after just about 10\,Myr; (2) the rapid infall of planets interior to the stellar corotation radius (initially at 0.025\,AU); and (3) the switch from the dynamical tide to the equilibrium tide regime at orbital frequencies $n=|2\Omega_\star|$ (initially at about 0.015\,AU). In panel (f), we compare a subset of these orbital evolution tracks to another set of tracks that we calculated using the conventional constant-$Q_\star$ model and assuming a constant stellar rotation period of 27\,d. In this model, the corotation radius is fixed at about 0.18\,AU and any planet interior to this relatively wide orbit will permanently fall into the star. Within 4.5 billion years, any hot Jupiter that started at 0.045\,AU to the star or closer is destructed. A pile-up is not reproduced however.

This discrepancy becomes even more apparent in the histograms shown in Figs.~\ref{fig:stelltid_hist}(a) and (b). Here we show snapshots of the simulated hot-Jupiter populations in the pure equilibrium tide model (a) and in the dynamical tide model with stellar evolution (b) at 10\,Myr (empty bars), 100\,Myr (striped bars), and 1\,000\,Myr (gray bars), respectively. While the equilibrium tide model suggests a steady removal of close-in planets over a billion years, the dynamical tide and stellar evolution model predicts that the hot-Jupiter population is essentially formed after between 10\, and 100\,Myr. Moreover, while the equilibrium tide model does not produce a pile-up, such a peak in the planet distribution occurs very naturally in the dynamical tide model with stellar evolution.

In Fig.~\ref{fig:stelltid_hist}(c) we investigate the effect of stellar metallicity on the pile-up efficiency and verify the finding of \citet{2017A&A...604A.113B} that more metal-rich young stars impose an enhanced tidal torque on their close-in massive planets. More explicitly, we find that the hot-Jupiter pile up is much more pronounced around metal-rich stars, increasing by about a factor of two as metallicity is increased from $Z=0.004$ (or ${\rm [Fe/H]}=-0.53$, gray shaded histogram) to $Z=0.02$ (or ${\rm [Fe/H]}=+0.28$, empty histogram).

Even within the early years of exoplanet observations it was becoming more apparent that stellar metallicity affects the likelihood of a star harboring a planet \citep{1997MNRAS.285..403G}. It was then debated whether the magnitude of this correlation may be different for different planetary masses, with larger planets possibly following a stronger correlation \citep{2014Natur.509..593B,2015ApJ...799L..26S}. Spectroscopy of {\it Kepler} exoplanet host stars delivered further evidence of a positive relation between stellar metallicity and hot-Jupiter occurrence rate \citep{2018AJ....155...89P}. 

\citet{2004ApJ...604..388I} argued that such a trend could originate in the protoplanetary disks since more-metal-rich disks forming metal-rich stars should also have had more solids available to form planets. Alternatively, recent simulations showed that stellar metallicity can also affect the tidally driven migration of close-in planets \citep{2017A&A...604A.113B}. Here we propose that the stronger tidal migration barrier around metal-rich stars shown in Fig.~\ref{fig:stelltid_hist}(c) can also explain the positive correlation between stellar metallicity and hot-Jupiter occurrence rate. The root of the correlation would then not be in the initial formation rate of hot Jupiters but in their survival rate against destruction via orbital migration toward to the star.

\section{Discussion}
\label{sec:discussion}

The very existence of a zero total torque location (the tidal migration barrier) is an important feature of our model and is rooted in the tidal dissipation implied by the stellar evolution tracks \citep{2016A&A...587A.105A,2017A&A...604A.113B}{. This dissipation is much higher and the resulting tidal dissipation factor $\bar{Q}_\star$ much smaller than previously assumed. For comparison, using canonical tidal dissipation factors of $10^5~{\leq}~\bar{Q}_\star~{\leq}~10^6$ \citep{
1998ApJ...500..428T,
2000ApJ...537L..61T,
2002ApJ...568L.117P,
2002A&A...394..241T,
2004ApJ...610..464D,
2004ApJ...614..955M,
2007ApJ...669.1298F,
2008IAUS..249..285Z,
2008ApJ...678.1396J,
2009ApJ...698.1357J,
2009ApJ...702.1413M,
2011A&A...528A...2B,
2012MNRAS.425.2567R,
2012ApJ...751..119B,
2014MNRAS.443.1451L}, we find that the location of zero torque would be between $0.008$\, and $0.012$\,AU (at orbital periods between 0.40\, and 0.66\,d), which is well inside the stellar corotation radius of the stellar evolution tracks and extremely close to the stellar surface. As a consequence, the lower stellar tidal dissipation assumed in previous studies cannot produce a tidal migration barrier in the type-II disk migration regime.

Our nominal disk properties come with significant uncertainties. For example, literature values of $\Sigma_{\rm p,0}$ span orders of magnitude, ranging from a few times $10\,{\rm g\,cm}^{-2}$ \citep{2004ApJ...606..520M} over some $100\,{\rm g\,cm}^{-2}$ \citep{2006A&A...445..747K,2013ApJ...779...59G,2017ApJ...835..230F} to $1\,000\,{\rm g\,cm}^{-2}$ and more at about 1\,AU \citep{1997ApJ...486..372B,2004ApJ...604..388I,2012ApJ...755...74K,2012ARA&A..50..211K}. Details depend on the dimensionality and assumptions of the respective models, and we chose $\Sigma_{\rm p,0}=1\,000\,{\rm g\,cm}^{-2}$ to reproduce the minimum-mass solar nebula for a viscous flaring disk with $\alpha=10^{-2}$. Different models would yield somewhat different estimates of the tidal migration barrier and, as a consequence, of the hot-Jupiter survival and formation rates.

In particular, our specific estimates of the tidal migration barrier rate around sun-like stars depend on our assumptions of the distribution of disk properties ($\log_{10}(\Sigma_{\rm p,0}/[{\rm g\,cm}^{-2}])~=~3~\pm~1$, $\log_{10}(\alpha)~=~-2~\pm~1$, and $\mu~=~1.85~\pm~0.55$) and of the efficiency of stellar dissipation ($\log_{10}(\langle\mathcal{D}\rangle_\omega)~=~-3.25\,~\pm~0.5$), which are free parameters in our model. We do have estimates for these quantities from simulations and observations, but detailed 3D magnetohydrodynamical simulations of the disk properties inside 0.1\,AU around accreting sun-like stars and of the evolution of tidal dissipation are required to validate or improve them.

In Eqs.~\eqref{eq:Gamma_II} and \eqref{eq:dot_a}, we have adopted a conventional description of the disk torque on the planet in the type-II migration regime, that is, we assume that the planet has separated the disk into an outer and an inner part with negligible flow between the two. Magnetic fields however could open up a magnetic cavity around the star, which would affect the disk torque and actually halt planet migration altogether \citep{1996Natur.380..606L,2002ApJ...574L..87K}. The critical distance for a cavity to form, referred to as the Alfv{\'e}n radius ($r_{\rm A}$), is determined by the equilibrium between the dynamic pressure of the disk matter and the magnetic pressure from the dipole field of the star. For T\,Tauri stars of about $0.8\,M_\odot$ and $2.5\,R_\odot$, the resulting value of $r_{\rm A}\approx0.05$\,AU \citep{2006ApJ...645L..73R} coincides equally well with the observed hot-Jupiter pile up as our theory of a tidal migration barrier. That said, 3D magnetohydrodynamic simulations showed that the mass inflow rate (and consequently $\Sigma$) interior to $r_{\rm A}$ can be significant. As a consequence, planet migration might in fact not stop near $r_{\rm A}$ \citep{2006ApJ...645L..73R} and an alternative mechanism would be required, which could be the tidal torque of the star, as we show.

The switch from positive to negative tidal torques at the stellar corotation radius, which is key to the tidal migration barrier, is only valid if both the eccentricity and the spin-orbit misalignment of the planet (its obliquity, $\psi_{\rm p}$) are small. Indeed, this switch does not apply for large obliquities, and in particular for $\psi_{\rm p}>90^\circ$ \citep{2009MNRAS.395.2268B,2018A&A...618A..90D}. In our calculations, the planet is assumed in the disk midplane and in the stellar equatorial plane. Many hot Jupiters however have actually been found in substantially misaligned orbits \citep{2012ApJ...757...18A} and tides actually might have played a key role in their formation \citep{2007ApJ...669.1298F,2014ApJ...790L..31D,2018MNRAS.480.1402A}. Several ways to put misaligned hot Jupiters in the context of this study could involve planet--planet gravitational interaction \citep{Naoz2011,Wu2011}, the Kozai-mechanism \citep{2008ApJ...678..498N}, or a combination of stellar tides, the disk torque, and high-eccentricity migration \citep{Wang2017} after an initial migration stranding at the tidal migration barrier. In fact, these would be compelling mechanisms to investigate in follow-up studies. Given our restriction to circular, non-oblique orbits, our model has limited applicability and in light of the various alternative mechanisms that could form close-in massive planets it might just be part of the solution of the formation of hot Jupiters.

In this paper we focus on the disk migration and stellar tidal interaction of hot Jupiters in the equatorial plane of their stars. A significant fraction of the hot Jupiters however exhibit substantial spin-orbit misalignments \citep{2015ARA&A..53..409W} and the corresponding tidal torques from their host stars differ significantly \citep{2012MNRAS.423..486L} from the zero-obliquity framework of this paper. Either these misalignments are created after the co-planar formation of hot Jupiters, for example through planet--planet scattering or the Lidov-Kozai mechanism, or the planets already formed in a warped or tilted protoplanetary disk. Since these processes act on different timescales and since the resulting stellar torque in oblique orbits is quite different from the formulation used in this paper, it might be possible that they can be observationally tested, although this treatment is beyond the scope of this study. For example, \citep{2011ApJ...729..138M} studied the outcomes of the spin-orbit evolution in the Lidov-Kozai migration scenario \citep{2007ApJ...669.1298F} and in the planet-scattering scenario \citep{2008ApJ...678..498N} and found evidence for the scattering model being the better explanation of the observed distribution of projected obliquities.

We have ignored in our model the effects of possible mass loss from the planet due to the Roche lobe overflow (RLO). The RLO initiates when the Roche lobe around the planet $R_{\rm R} \approx a (M_{\rm p}/M_{\rm p})^{1/3}$ becomes as small as the planetary radius. Even in the case of a highly inflated, young gas giant with a radius of $2\,R_{\rm J}$, RLO at $R_{\rm R}\approx2\,R_{\rm J}$ will only be reached at a semimajor axis $a\approx0.01$\,AU, that is, very close to the stellar surface, which is at 0.0093\,AU for a 1\,Myr young solar type star. Hence, there will be a very sensitive fine tuning involved between extreme planetary inflation ($R_{\rm p}>2\,R_{\rm J}$), a sufficiently small star ($R_\star<2\,R_\odot$), and sufficiently weak tidally driven orbital decay to prevent an inward spiralling hot Jupiter from falling into the star by RLO. Some of the known planets might indeed be or have been affected by RLO \citep{2016CeMDA.126..227J} and the inclusion of RLO could be relevant interior to the inner cavity of our model. The modeling of RLO however requires some parameterization (or even modeling of) planetary structure, which is beyond the scope of this paper.

As for the stellar spin-up, in our scenario of planet migration under the effect of stellar tides, hot Jupiters would either fall into their star within the first $\approx 10$\,Myr of their lifetime if the tidal torque is too weak to stop migration. Or they would stop beyond the stellar corotation radius at the tidal migration barrier, where they would act to slow down the stellar rotation. In our picture, hot Jupiters spin up their stars if they fall through the corotation radius and get swallowed by the star, or they survive and act as a rotational brake during the first $\approx 100$\,Myr of the lifetime of the host star.

Tidally excited waves in the stellar radiation zones \citep{1998ApJ...507..938G,2007ApJ...661.1180O,2017MNRAS.470.2054C} or the wave breaking mechanism at the stellar center \citep{2010MNRAS.404.1849B} could trigger additional tidal dissipation beyond the processes in the convective envelope considered in this study. Further refinements of this theory could be achieved through the consistent modeling of the radial profile of the stellar density, the latter of which is assumed to be constant (though different) in both the stellar core and the envelope in our model \citep{2013MNRAS.429..613O}.

We have verified the findings of \citet{2017A&A...604A.113B} that more metal-rich stars exert a stronger tidal torque on their close-in hot-Jupiter companions. This result could at least partly explain the finding of a lower abundance of massive planets ($M_{\rm p}>4\,M_{\rm Jup}$) in short-period orbits ($<10$\,d) around low-metallicity stars by \citet{2018AJ....156..221N}. In other words, this observed low abundance could be due to the lower tidal dissipation rates in young, solar-mass low-metallicity stars and the resulting lower migration survival rates of the planets.

\section{Conclusions}

We present a new model for the formation of hot Jupiters under the combined effects of the dynamical tide in the convective envelope of the star and type-II planet migration. First, we calculate the nominal tidal torques of a young, highly dissipative solar-type star and of a 2D viscous disk in thermal equilibrium with radial temperature, gas surface density, and viscosity dependences acting on a Jupiter-sized migrating planet. We study a system made up of a $2\,R_\odot$ star with a range of possible spin periods according to observations of young star clusters, and a disk gas surface density similar to the minimum-mass solar nebula ($\Sigma_{\rm p}=1\,000\,{\rm g\,cm}^{-2}\,(a/{\rm AU})^{-3/2}$, $X=0.7$, $Y=0.28$), with different $\alpha$ viscosity values ($10^{-1}, 10^{-2}, 10^{-3}$) and a disk opacity of $\kappa=10^{-7}\,{\rm m}^2\,{\rm kg}^{-1}$. We also consider two possible scenarios for the radial position of the inner disk magnetic cavity. In one scenario the location is randomly drawn from a normal distribution as per $r_{\rm mc}=0.05\,(\pm0.015)$\,AU, while in the other scenario it is coupled to the stellar rotation period and is located at the corotation radius.

Our Monte Carlo simulations of several times 10\,000 realizations of this parameterized star--planet--disk model suggest that hot Jupiters that form and migrate towards their sun-like host stars survive inwards migration in about 2.9\, to 15.6\,\% of all cases. This result is almost unaffected by the choice of the scenario for the location of the magnetic cavity. In combination with the bias-corrected observed hot-Jupiter occurrence rate of $1.2\,\pm0.38\,\%$ \citep{2012ApJ...753..160W}, this implies an initial hot-Jupiter formation rate of less than between 41\, and 8\,\% around sun-like stars. In other words, we predict that between $<8$\,\% and $<41$\,\% of all sun-like stars initially give birth to a hot Jupiter because our simulations show that between about 2.9\, and 15.6\,\% of these planets ultimately survive disk migration near the tidal migration barrier. This would explain the observed hot-Jupiter occurrence of about one planet around every 100 sun-like stars ($1.2\,\%$). The ``$<$'' sign here refers to an upper limit of hot-Jupiter formation through the formation channel described in this paper and it absorbs hot-Jupiter formation through any other formation channel.

We then consider a second evolutionary stage of the system, in which the protoplanetary nebula has been fully accreted onto the star and in which the planetary orbit evolves under the effects of the stellar tide only. We couple the differential equation for the orbital migration of the planet with precomputed stellar evolution tracks for three different stellar metallicities, which take into account the the internal evolution of the star and therefore the long-term weakening of tidal dissipation on a timescale of one billion years. These orbital simulations naturally produce a pile-up of planets near 0.05\,AU, which is similar to that observed in the hot-Jupiter population.

In our hot-Jupiter formation model, the protoplanets either fall into their host star or they reach their tidal migration barrier within the first ${\approx}10$\,Myr of the lifetime of the system. The fraction of hot Jupiters that survives inward migration beyond the stellar corotation radius (2.9\,\% - 15.6\,\% in our simulations) would then be repelled by the star as the disk (and therefore the disk torque) is being removed, implying a first-in-then-out migration scenario for many of the hot Jupiters observed near 0.05\,AU today. The ultimate fate of the planet is determined by the combined effects of the negative disk torque (here in the type-II migration regime) and the tidal torque of the star acting on the planet, the latter being positive beyond the stellar corotation radius. This timescale for the formation of the pile-up is much shorter than the previously predicted $500$\,Myr \citep{2004ApJ...610..464D}. 

Hot Jupiters found at about 0.05\,AU today were beyond the corotation radius when the star was tidally highly dissipative, and so they have been pushed away from the star once the protoplanetary disk was accreted onto it. Nowadays, these planets are usually within the corotation radius of their billion-year-old stars but tidal dissipation of the dynamical tide is now extremely weak and does not lead to significant orbital decay. This is in agreement with the null detection of tidally driven orbital decay observed for the hot Jupiters WASP-43\,b \citep{2016AJ....151..137H}, OGLE-TR-113\,b \citep{2016MNRAS.455.1334H}, and WASP-46\,b \citep{2018MNRAS.473.5126P}, which suggest $Q_\star~{\gg}~10^5$, and with the high tidal quality factors found in a recent census of the hot-Jupiter population \citep{2018MNRAS.476.2542C}. Generally speaking, we predict that the tidally driven orbital decay of hot Jupiters around Sun-like main sequence stars cannot be observed in most cases due to the extremely ineffective tidal dissipation in the convective envelope, producing frequency-averaged tidal dissipation factors of $10^8~{\lesssim}~\bar{Q}_\star~{\lesssim}~10^9$. Another consequence of our simulations is that hot Jupiters do not tend to spin-up their host star if they are stranded beyond the corotation radius of the star during the phase of planet migration. In the long term, they  also do not significantly slow down the magnetic braking of the star. For an in-depth treatment of tidal evolution under the effects of an evolving stellar wind and magnetic braking see \citep{2019A&A...621A.124B}.

The orbital decay interpretation for the observed transit timing variation of WASP-12\,b \citep{2016A&A...588L...6M} is a peculiar case, the star being the largest ($1.57\,\pm0.07\,R_\odot$) and most massive ($1.35\,\pm0.14\,M_\odot$) to host a hot Jupiter within 0.025\,AU \citep{2009ApJ...693.1920H}. Hence, the tidal decay interpretation might actually be valid and be caused by additional tidally dissipative effects in the stellar radiative core that are not taken into account in our model \citep{2017ApJ...849L..11W}. Alternatively, obliquity tides in this highly inflated planet could drive its orbital decay, which would require the additional torques from a second, roughly Neptune-mass planet within 0.04\,AU of the star \citep{2018ApJ...869L..15M}. Either way, WASP-12\,b  is certainly a benchmark object to test dynamical tide theory.

Our model also explains the observed pile-up of hot Jupiters in RV surveys as an outcome of tidally driven planet migration on a 10 - 100\,Myr timescale, when stellar tidal dissipation is still highly efficient and the star is still a fast rotator with a close-in corotation radius. Our numerical orbital simulations show that any hot Jupiters that survived disk migration naturally accumulate at about 0.05\,AU.

Our numerical orbital simulations, which are coupled to stellar evolution tracks of different stellar metallicities, show that the tidal migration barrier is more effective for planets around metal-rich stars. This could explain why increasingly metal-rich stars have more hot Jupiters \citep{2018AJ....155...89P}: these planets might have met a more efficient tidal migration barrier imposed by the stars during the early phase of planet migration.

\begin{acknowledgements}
The author thanks Emeline Bolmont and Cilia Damiani for insightful discussions about the theory of stellar tidal friction, Willy Kley for comments on the disk model, Florian Gallet for advice on the precomputed stellar evolution tracks, Brian Jackson for feedback on a draft version of this manuscript, David Ciardi from the ExoFOP team for technical support on using the NASA Exoplanet Archive, John Southworth for help with the TEPCAT online catalog, and an anonymous referee for comments that improved the clarity of the paper. The author made use of NASA's ADS Bibliographic Services. Computations have been performed with {\tt ipython 4.0.1} on {\tt python 2.7.10} \citep{PER-GRA:2007} and with {\tt gnuplot 5.2} \citep{gnuplot}. This work was supported by the German space agency (Deutsches Zentrum f\"ur Luft- und Raumfahrt) under PLATO Data Center grant 50OO1501.
\end{acknowledgements}

\bibliographystyle{aa}
\bibliography{ms}

\begin{thebibliography}{106}
\expandafter\ifx\csname natexlab\endcsname\relax\def\natexlab#1{#1}\fi

\bibitem[{{Albrecht} {et~al.}(2012){Albrecht}, {Winn}, {Johnson}, {Howard},
  {Marcy}, {Butler}, {Arriagada}, {Crane}, {Shectman}, {Thompson}, {Hirano},
  {Bakos}, \& {Hartman}}]{2012ApJ...757...18A}
{Albrecht}, S., {Winn}, J.~N., {Johnson}, J.~A., {et~al.} 2012, \apj, 757, 18

\bibitem[{{Alibert} {et~al.}(2005){Alibert}, {Mordasini}, {Benz}, \&
  {Winisdoerffer}}]{2005A&A...434..343A}
{Alibert}, Y., {Mordasini}, C., {Benz}, W., \& {Winisdoerffer}, C. 2005, \aap,
  434, 343

\bibitem[{{Amard} {et~al.}(2016){Amard}, {Palacios}, {Charbonnel}, {Gallet}, \&
  {Bouvier}}]{2016A&A...587A.105A}
{Amard}, L., {Palacios}, A., {Charbonnel}, C., {Gallet}, F., \& {Bouvier}, J.
  2016, \aap, 587, A105

\bibitem[{{Anderson} \& {Lai}(2018)}]{2018MNRAS.480.1402A}
{Anderson}, K.~R. \& {Lai}, D. 2018, \mnras, 480, 1402

\bibitem[{{Barker} \& {Ogilvie}(2009)}]{2009MNRAS.395.2268B}
{Barker}, A.~J. \& {Ogilvie}, G.~I. 2009, \mnras, 395, 2268

\bibitem[{{Barker} \& {Ogilvie}(2010)}]{2010MNRAS.404.1849B}
{Barker}, A.~J. \& {Ogilvie}, G.~I. 2010, \mnras, 404, 1849

\bibitem[{{Beaug{\'e}} \& {Nesvorn{\'y}}(2012)}]{2012ApJ...751..119B}
{Beaug{\'e}}, C. \& {Nesvorn{\'y}}, D. 2012, \apj, 751, 119

\bibitem[{{Bell} {et~al.}(1997){Bell}, {Cassen}, {Klahr}, \&
  {Henning}}]{1997ApJ...486..372B}
{Bell}, K.~R., {Cassen}, P.~M., {Klahr}, H.~H., \& {Henning}, T. 1997, \apj,
  486, 372

\bibitem[{{Benbakoura} {et~al.}(2019){Benbakoura}, {R{\'e}ville}, {Brun}, {Le
  Poncin-Lafitte}, \& {Mathis}}]{2019A&A...621A.124B}
{Benbakoura}, M., {R{\'e}ville}, V., {Brun}, A.~S., {Le Poncin-Lafitte}, C., \&
  {Mathis}, S. 2019, \aap, 621, A124

\bibitem[{{Ben{\'{\i}}tez-Llambay} {et~al.}(2011){Ben{\'{\i}}tez-Llambay},
  {Masset}, \& {Beaug{\'e}}}]{2011A&A...528A...2B}
{Ben{\'{\i}}tez-Llambay}, P., {Masset}, F., \& {Beaug{\'e}}, C. 2011, \aap,
  528, A2

\bibitem[{{Bolmont} {et~al.}(2017){Bolmont}, {Gallet}, {Mathis}, {Charbonnel},
  {Amard}, \& {Alibert}}]{2017A&A...604A.113B}
{Bolmont}, E., {Gallet}, F., {Mathis}, S., {et~al.} 2017, \aap, 604, A113

\bibitem[{{Bolmont} \& {Mathis}(2016)}]{2016CeMDA.126..275B}
{Bolmont}, E. \& {Mathis}, S. 2016, Celestial Mechanics and Dynamical
  Astronomy, 126, 275

\bibitem[{{Bolmont} {et~al.}(2012){Bolmont}, {Raymond}, {Leconte}, \&
  {Matt}}]{2012A&A...544A.124B}
{Bolmont}, E., {Raymond}, S.~N., {Leconte}, J., \& {Matt}, S.~P. 2012, \aap,
  544, A124

\bibitem[{{Bouvier} {et~al.}(1997){Bouvier}, {Forestini}, \&
  {Allain}}]{1997A&A...326.1023B}
{Bouvier}, J., {Forestini}, M., \& {Allain}, S. 1997, \aap, 326, 1023

\bibitem[{{Buchhave} {et~al.}(2014){Buchhave}, {Bizzarro}, {Latham},
  {Sasselov}, {Cochran}, {Endl}, {Isaacson}, {Juncher}, \&
  {Marcy}}]{2014Natur.509..593B}
{Buchhave}, L.~A., {Bizzarro}, M., {Latham}, D.~W., {et~al.} 2014, \nat, 509,
  593

\bibitem[{{Chernov} {et~al.}(2017){Chernov}, {Ivanov}, \&
  {Papaloizou}}]{2017MNRAS.470.2054C}
{Chernov}, S.~V., {Ivanov}, P.~B., \& {Papaloizou}, J.~C.~B. 2017, \mnras, 470,
  2054

\bibitem[{{Collier Cameron} \& {Jardine}(2018)}]{2018MNRAS.476.2542C}
{Collier Cameron}, A. \& {Jardine}, M. 2018, \mnras, 476, 2542

\bibitem[{{Damiani} \& {Mathis}(2018)}]{2018A&A...618A..90D}
{Damiani}, C. \& {Mathis}, S. 2018, \aap, 618, A90

\bibitem[{{D'Angelo} \& {Bodenheimer}(2013)}]{2013ApJ...778...77D}
{D'Angelo}, G. \& {Bodenheimer}, P. 2013, \apj, 778, 77

\bibitem[{{Dawson}(2014)}]{2014ApJ...790L..31D}
{Dawson}, R.~I. 2014, \apjl, 790, L31

\bibitem[{{Dawson} \& {Johnson}(2018)}]{2018ARA&A..56..175D}
{Dawson}, R.~I. \& {Johnson}, J.~A. 2018, \araa, 56, 175

\bibitem[{{Dobbs-Dixon} {et~al.}(2004){Dobbs-Dixon}, {Lin}, \&
  {Mardling}}]{2004ApJ...610..464D}
{Dobbs-Dixon}, I., {Lin}, D.~N.~C., \& {Mardling}, R.~A. 2004, \apj, 610, 464

\bibitem[{{Duffell} {et~al.}(2014){Duffell}, {Haiman}, {MacFadyen}, {D'Orazio},
  \& {Farris}}]{2014ApJ...792L..10D}
{Duffell}, P.~C., {Haiman}, Z., {MacFadyen}, A.~I., {D'Orazio}, D.~J., \&
  {Farris}, B.~D. 2014, \apjl, 792, L10

\bibitem[{{D{\"u}rmann} \& {Kley}(2015)}]{2015A&A...574A..52D}
{D{\"u}rmann}, C. \& {Kley}, W. 2015, \aap, 574, A52

\bibitem[{Efroimsky \& Makarov(2013)}]{Efroimsky2013}
Efroimsky, M. \& Makarov, V.~V. 2013, \apj, 764, 26

\bibitem[{{Eggleton} {et~al.}(1998){Eggleton}, {Kiseleva}, \&
  {Hut}}]{1998ApJ...499..853E}
{Eggleton}, P.~P., {Kiseleva}, L.~G., \& {Hut}, P. 1998, \apj, 499, 853

\bibitem[{{Fabrycky} \& {Tremaine}(2007)}]{2007ApJ...669.1298F}
{Fabrycky}, D. \& {Tremaine}, S. 2007, \apj, 669, 1298

\bibitem[{{Flock} {et~al.}(2017){Flock}, {Fromang}, {Turner}, \&
  {Benisty}}]{2017ApJ...835..230F}
{Flock}, M., {Fromang}, S., {Turner}, N.~J., \& {Benisty}, M. 2017, \apj, 835,
  230

\bibitem[{{Goldreich} \& {Tremaine}(1979)}]{1979ApJ...233..857G}
{Goldreich}, P. \& {Tremaine}, S. 1979, \apj, 233, 857

\bibitem[{{Gonzalez}(1997)}]{1997MNRAS.285..403G}
{Gonzalez}, G. 1997, \mnras, 285, 403

\bibitem[{{Goodman} \& {Dickson}(1998)}]{1998ApJ...507..938G}
{Goodman}, J. \& {Dickson}, E.~S. 1998, \apj, 507, 938

\bibitem[{{Greenberg}(2009)}]{2009ApJ...698L..42G}
{Greenberg}, R. 2009, \apjl, 698, L42

\bibitem[{{Gressel} {et~al.}(2013){Gressel}, {Nelson}, {Turner}, \&
  {Ziegler}}]{2013ApJ...779...59G}
{Gressel}, O., {Nelson}, R.~P., {Turner}, N.~J., \& {Ziegler}, U. 2013, \apj,
  779, 59

\bibitem[{{Guo} {et~al.}(2017){Guo}, {Johnson}, {Mann}, {Kraus}, {Curtis}, \&
  {Latham}}]{2017ApJ...838...25G}
{Guo}, X., {Johnson}, J.~A., {Mann}, A.~W., {et~al.} 2017, \apj, 838, 25

\bibitem[{{Haisch} {et~al.}(2001){Haisch}, {Lada}, \&
  {Lada}}]{2001ApJ...553L.153H}
{Haisch}, Jr., K.~E., {Lada}, E.~A., \& {Lada}, C.~J. 2001, \apjl, 553, L153

\bibitem[{{Hansen}(2010)}]{2010ApJ...723..285H}
{Hansen}, B.~M.~S. 2010, \apj, 723, 285

\bibitem[{{Hansen}(2012)}]{2012ApJ...757....6H}
{Hansen}, B.~M.~S. 2012, \apj, 757, 6

\bibitem[{Hasegawa \& Pudritz(2010)}]{Hasegawa2010}
Hasegawa, Y. \& Pudritz, R.~E. 2010, \apjl, 710, L167

\bibitem[{{Hayashi}(1981)}]{1981PThPS..70...35H}
{Hayashi}, C. 1981, Progress of Theoretical Physics Supplement, 70, 35

\bibitem[{{Hebb} {et~al.}(2009){Hebb}, {Collier-Cameron}, {Loeillet},
  {Pollacco}, {H{\'e}brard}, {Street}, {Bouchy}, {Stempels}, {Moutou},
  {Simpson}, {Udry}, {Joshi}, {West}, {Skillen}, {Wilson}, {McDonald},
  {Gibson}, {Aigrain}, {Anderson}, {Benn}, {Christian}, {Enoch}, {Haswell},
  {Hellier}, {Horne}, {Irwin}, {Lister}, {Maxted}, {Mayor}, {Norton}, {Parley},
  {Pont}, {Queloz}, {Smalley}, \& {Wheatley}}]{2009ApJ...693.1920H}
{Hebb}, L., {Collier-Cameron}, A., {Loeillet}, B., {et~al.} 2009, \apj, 693,
  1920

\bibitem[{{Heller}(2018)}]{2018ascl.soft06014H}
{Heller}, R. 2018, {pile-up: Monte Carlo simulations of star-disk torques on
  hot Jupiters}, Astrophysics Source Code Library

\bibitem[{{Henning} \& {Stognienko}(1996)}]{1996A&A...311..291H}
{Henning}, T. \& {Stognienko}, R. 1996, \aap, 311, 291

\bibitem[{{Hoyer} {et~al.}(2016{\natexlab{a}}){Hoyer}, {L{\'o}pez-Morales},
  {Rojo}, {Minniti}, \& {Adams}}]{2016MNRAS.455.1334H}
{Hoyer}, S., {L{\'o}pez-Morales}, M., {Rojo}, P., {Minniti}, D., \& {Adams},
  E.~R. 2016{\natexlab{a}}, \mnras, 455, 1334

\bibitem[{{Hoyer} {et~al.}(2016{\natexlab{b}}){Hoyer}, {Pall{\'e}}, {Dragomir},
  \& {Murgas}}]{2016AJ....151..137H}
{Hoyer}, S., {Pall{\'e}}, E., {Dragomir}, D., \& {Murgas}, F.
  2016{\natexlab{b}}, \aj, 151, 137

\bibitem[{{Hubeny}(1990)}]{1990ApJ...351..632H}
{Hubeny}, I. 1990, \apj, 351, 632

\bibitem[{{Hut}(1981)}]{1981A&A....99..126H}
{Hut}, P. 1981, \aap, 99, 126

\bibitem[{{Ida} \& {Lin}(2004)}]{2004ApJ...604..388I}
{Ida}, S. \& {Lin}, D.~N.~C. 2004, \apj, 604, 388

\bibitem[{{Irwin} {et~al.}(2008){Irwin}, {Hodgkin}, {Aigrain}, {Bouvier},
  {Hebb}, {Irwin}, \& {Moraux}}]{2008MNRAS.384..675I}
{Irwin}, J., {Hodgkin}, S., {Aigrain}, S., {et~al.} 2008, \mnras, 384, 675

\bibitem[{{Jackson} {et~al.}(2009){Jackson}, {Barnes}, \&
  {Greenberg}}]{2009ApJ...698.1357J}
{Jackson}, B., {Barnes}, R., \& {Greenberg}, R. 2009, \apj, 698, 1357

\bibitem[{{Jackson} {et~al.}(2008){Jackson}, {Greenberg}, \&
  {Barnes}}]{2008ApJ...678.1396J}
{Jackson}, B., {Greenberg}, R., \& {Barnes}, R. 2008, \apj, 678, 1396

\bibitem[{{Jackson} {et~al.}(2016){Jackson}, {Jensen}, {Peacock}, {Arras}, \&
  {Penev}}]{2016CeMDA.126..227J}
{Jackson}, B., {Jensen}, E., {Peacock}, S., {Arras}, P., \& {Penev}, K. 2016,
  Celestial Mechanics and Dynamical Astronomy, 126, 227

\bibitem[{{King} {et~al.}(2007){King}, {Pringle}, \&
  {Livio}}]{2007MNRAS.376.1740K}
{King}, A.~R., {Pringle}, J.~E., \& {Livio}, M. 2007, \mnras, 376, 1740

\bibitem[{{Klahr} \& {Kley}(2006)}]{2006A&A...445..747K}
{Klahr}, H. \& {Kley}, W. 2006, \aap, 445, 747

\bibitem[{{Kley} \& {Nelson}(2012)}]{2012ARA&A..50..211K}
{Kley}, W. \& {Nelson}, R.~P. 2012, \araa, 50, 211

\bibitem[{{Kretke} \& {Lin}(2012)}]{2012ApJ...755...74K}
{Kretke}, K.~A. \& {Lin}, D.~N.~C. 2012, \apj, 755, 74

\bibitem[{{Kuchner} \& {Lecar}(2002)}]{2002ApJ...574L..87K}
{Kuchner}, M.~J. \& {Lecar}, M. 2002, \apjl, 574, L87

\bibitem[{{Lagarde} {et~al.}(2012){Lagarde}, {Decressin}, {Charbonnel},
  {Eggenberger}, {Ekstr{\"o}m}, \& {Palacios}}]{2012A&A...543A.108L}
{Lagarde}, N., {Decressin}, T., {Charbonnel}, C., {et~al.} 2012, \aap, 543,
  A108

\bibitem[{{Lai}(2012)}]{2012MNRAS.423..486L}
{Lai}, D. 2012, \mnras, 423, 486

\bibitem[{{Lanza} \& {Shkolnik}(2014)}]{2014MNRAS.443.1451L}
{Lanza}, A.~F. \& {Shkolnik}, E.~L. 2014, \mnras, 443, 1451

\bibitem[{{Leconte} {et~al.}(2010){Leconte}, {Chabrier}, {Baraffe}, \&
  {Levrard}}]{2010A&A...516A..64L}
{Leconte}, J., {Chabrier}, G., {Baraffe}, I., \& {Levrard}, B. 2010, \aap, 516,
  A64

\bibitem[{{Lin} {et~al.}(1996){Lin}, {Bodenheimer}, \&
  {Richardson}}]{1996Natur.380..606L}
{Lin}, D.~N.~C., {Bodenheimer}, P., \& {Richardson}, D.~C. 1996, \nat, 380, 606

\bibitem[{{Lin} \& {Papaloizou}(1986)}]{1986ApJ...309..846L}
{Lin}, D.~N.~C. \& {Papaloizou}, J. 1986, \apj, 309, 846

\bibitem[{{Maciejewski} {et~al.}(2016){Maciejewski}, {Dimitrov},
  {Fern{\'a}ndez}, {Sota}, {Nowak}, {Ohlert}, {Nikolov}, {Bukowiecki}, {Hinse},
  {Pall{\'e}}, {Tingley}, {Kjurkchieva}, {Lee}, \& {Lee}}]{2016A&A...588L...6M}
{Maciejewski}, G., {Dimitrov}, D., {Fern{\'a}ndez}, M., {et~al.} 2016, \aap,
  588, L6

\bibitem[{{Mardling} \& {Lin}(2004)}]{2004ApJ...614..955M}
{Mardling}, R.~A. \& {Lin}, D.~N.~C. 2004, \apj, 614, 955

\bibitem[{{Martin} {et~al.}(2019){Martin}, {Nixon}, {Pringle}, \&
  {Livio}}]{2019NewA...70....7M}
{Martin}, R.~G., {Nixon}, C.~J., {Pringle}, J.~E., \& {Livio}, M. 2019, \na,
  70, 7

\bibitem[{{Mathis}(2015)}]{Mathis2015}
{Mathis}, S. 2015, \aap, 580, L3

\bibitem[{{Mayor} \& {Queloz}(1995)}]{1995Natur.378..355M}
{Mayor}, M. \& {Queloz}, D. 1995, \nat, 378, 355

\bibitem[{{Menou} \& {Goodman}(2004)}]{2004ApJ...606..520M}
{Menou}, K. \& {Goodman}, J. 2004, \apj, 606, 520

\bibitem[{{Miller} {et~al.}(2009){Miller}, {Fortney}, \&
  {Jackson}}]{2009ApJ...702.1413M}
{Miller}, N., {Fortney}, J.~J., \& {Jackson}, B. 2009, \apj, 702, 1413

\bibitem[{{Millholland} \& {Laughlin}(2018)}]{2018ApJ...869L..15M}
{Millholland}, S. \& {Laughlin}, G. 2018, \apjl, 869, L15

\bibitem[{{Morton} \& {Johnson}(2011)}]{2011ApJ...729..138M}
{Morton}, T.~D. \& {Johnson}, J.~A. 2011, \apj, 729, 138

\bibitem[{{Murray} \& {Dermott}(1999)}]{murray99}
{Murray}, C.~D. \& {Dermott}, S.~F. 1999, {Solar System Dynamics} (Cambridge
  University Press)

\bibitem[{{Nagasawa} {et~al.}(2008){Nagasawa}, {Ida}, \&
  {Bessho}}]{2008ApJ...678..498N}
{Nagasawa}, M., {Ida}, S., \& {Bessho}, T. 2008, \apj, 678, 498

\bibitem[{Naoz {et~al.}(2011)Naoz, Farr, Lithwick, Rasio, \&
  Teyssandier}]{Naoz2011}
Naoz, S., Farr, W.~M., Lithwick, Y., Rasio, F.~A., \& Teyssandier, J. 2011,
  Nature, 473, 187 EP

\bibitem[{{Narang} {et~al.}(2018){Narang}, {Manoj}, {Furlan}, {Mordasini},
  {Henning}, {Mathew}, {Banyal}, \& {Sivarani}}]{2018AJ....156..221N}
{Narang}, M., {Manoj}, P., {Furlan}, E., {et~al.} 2018, \aj, 156, 221

\bibitem[{{Nelson} {et~al.}(2000){Nelson}, {Papaloizou}, {Masset}, \&
  {Kley}}]{2000MNRAS.318...18N}
{Nelson}, R., {Papaloizou}, J.~C.~B., {Masset}, F., \& {Kley}, W. 2000, \mnras,
  318, 18

\bibitem[{{Nordstr{\"o}m} {et~al.}(2004){Nordstr{\"o}m}, {Mayor}, {Andersen},
  {Holmberg}, {Pont}, {J{\o}rgensen}, {Olsen}, {Udry}, \&
  {Mowlavi}}]{2004A&A...418..989N}
{Nordstr{\"o}m}, B., {Mayor}, M., {Andersen}, J., {et~al.} 2004, \aap, 418, 989

\bibitem[{{Ogilvie}(2013)}]{2013MNRAS.429..613O}
{Ogilvie}, G.~I. 2013, \mnras, 429, 613

\bibitem[{{Ogilvie} \& {Lin}(2007)}]{2007ApJ...661.1180O}
{Ogilvie}, G.~I. \& {Lin}, D.~N.~C. 2007, \apj, 661, 1180

\bibitem[{{P{\"a}tzold} \& {Rauer}(2002)}]{2002ApJ...568L.117P}
{P{\"a}tzold}, M. \& {Rauer}, H. 2002, \apjl, 568, L117

\bibitem[{P\'erez \& Granger(2007)}]{PER-GRA:2007}
P\'erez, F. \& Granger, B.~E. 2007, {C}omput. {S}ci. {E}ng., 9, 21

\bibitem[{{Petigura} {et~al.}(2018){Petigura}, {Marcy}, {Winn}, {Weiss},
  {Fulton}, {Howard}, {Sinukoff}, {Isaacson}, {Morton}, \&
  {Johnson}}]{2018AJ....155...89P}
{Petigura}, E.~A., {Marcy}, G.~W., {Winn}, J.~N., {et~al.} 2018, \aj, 155, 89

\bibitem[{{Petrucci} {et~al.}(2018){Petrucci}, {Jofr{\'e}}, {Ferrero},
  {C{\'u}neo}, {Saker}, {Lovos}, {G{\'o}mez}, \& {Mauas}}]{2018MNRAS.473.5126P}
{Petrucci}, R., {Jofr{\'e}}, E., {Ferrero}, L.~V., {et~al.} 2018, \mnras, 473,
  5126

\bibitem[{{Plavchan} \& {Bilinski}(2013)}]{2013ApJ...769...86P}
{Plavchan}, P. \& {Bilinski}, C. 2013, \apj, 769, 86

\bibitem[{{Pollack} {et~al.}(1994){Pollack}, {Hollenbach}, {Beckwith},
  {Simonelli}, {Roush}, \& {Fong}}]{1994ApJ...421..615P}
{Pollack}, J.~B., {Hollenbach}, D., {Beckwith}, S., {et~al.} 1994, \apj, 421,
  615

\bibitem[{{Rao} {et~al.}(2018){Rao}, {Meynet}, {Eggenberger}, {Haemmerl{\'e}},
  {Privitera}, {Georgy}, {Ekstr{\"o}m}, \& {Mordasini}}]{2018A&A...618A..18R}
{Rao}, S., {Meynet}, G., {Eggenberger}, P., {et~al.} 2018, \aap, 618, A18

\bibitem[{{Rice} {et~al.}(2012){Rice}, {Veljanoski}, \& {Collier
  Cameron}}]{2012MNRAS.425.2567R}
{Rice}, W.~K.~M., {Veljanoski}, J., \& {Collier Cameron}, A. 2012, \mnras, 425,
  2567

\bibitem[{{Romanova} \& {Lovelace}(2006)}]{2006ApJ...645L..73R}
{Romanova}, M.~M. \& {Lovelace}, R.~V.~E. 2006, \apjl, 645, L73

\bibitem[{{Schlaufman}(2015)}]{2015ApJ...799L..26S}
{Schlaufman}, K.~C. 2015, \apjl, 799, L26

\bibitem[{{Schneider} {et~al.}(2011){Schneider}, {Dedieu}, {Le Sidaner},
  {Savalle}, \& {Zolotukhin}}]{2011A&A...532A..79S}
{Schneider}, J., {Dedieu}, C., {Le Sidaner}, P., {Savalle}, R., \&
  {Zolotukhin}, I. 2011, \aap, 532, A79

\bibitem[{{Shakura} \& {Sunyaev}(1973)}]{1973A&A....24..337S}
{Shakura}, N.~I. \& {Sunyaev}, R.~A. 1973, \aap, 24, 337

\bibitem[{{Southworth}(2011)}]{2011MNRAS.417.2166S}
{Southworth}, J. 2011, \mnras, 417, 2166

\bibitem[{{Stassun} {et~al.}(1999){Stassun}, {Mathieu}, {Mazeh}, \&
  {Vrba}}]{1999AJ....117.2941S}
{Stassun}, K.~G., {Mathieu}, R.~D., {Mazeh}, T., \& {Vrba}, F.~J. 1999, \aj,
  117, 2941

\bibitem[{{Terquem}(2003)}]{2003MNRAS.341.1157T}
{Terquem}, C.~E.~J.~M.~L.~J. 2003, \mnras, 341, 1157

\bibitem[{{Trilling}(2000)}]{2000ApJ...537L..61T}
{Trilling}, D.~E. 2000, \apjl, 537, L61

\bibitem[{{Trilling} {et~al.}(1998){Trilling}, {Benz}, {Guillot}, {Lunine},
  {Hubbard}, \& {Burrows}}]{1998ApJ...500..428T}
{Trilling}, D.~E., {Benz}, W., {Guillot}, T., {et~al.} 1998, \apj, 500, 428

\bibitem[{{Trilling} {et~al.}(2002){Trilling}, {Lunine}, \&
  {Benz}}]{2002A&A...394..241T}
{Trilling}, D.~E., {Lunine}, J.~I., \& {Benz}, W. 2002, \aap, 394, 241

\bibitem[{Wang {et~al.}(2017)Wang, lin Zhou, hui gen, \& Meng}]{Wang2017}
Wang, Y., lin Zhou, J., hui gen, L., \& Meng, Z. 2017, \apj, 848, 20

\bibitem[{{Ward}(1997)}]{1997Icar..126..261W}
{Ward}, W.~R. 1997, \icarus, 126, 261

\bibitem[{{Weinberg} {et~al.}(2017){Weinberg}, {Sun}, {Arras}, \&
  {Essick}}]{2017ApJ...849L..11W}
{Weinberg}, N.~N., {Sun}, M., {Arras}, P., \& {Essick}, R. 2017, \apjl, 849,
  L11

\bibitem[{Williams \& Kelley(2015)}]{gnuplot}
Williams, T. \& Kelley, C. 2015, Gnuplot 5.0 Reference Manual (Samurai Media
  Limited)

\bibitem[{{Winn} \& {Fabrycky}(2015)}]{2015ARA&A..53..409W}
{Winn}, J.~N. \& {Fabrycky}, D.~C. 2015, \araa, 53, 409

\bibitem[{{Wright} {et~al.}(2012){Wright}, {Marcy}, {Howard}, {Johnson},
  {Morton}, \& {Fischer}}]{2012ApJ...753..160W}
{Wright}, J.~T., {Marcy}, G.~W., {Howard}, A.~W., {et~al.} 2012, \apj, 753, 160

\bibitem[{Wu \& Lithwick(2011)}]{Wu2011}
Wu, Y. \& Lithwick, Y. 2011, \apj, 735, 109

\bibitem[{{Zahn}(1977)}]{1977A&A....57..383Z}
{Zahn}, J.-P. 1977, \aap, 57, 383

\bibitem[{{Zhou} \& {Lin}(2008)}]{2008IAUS..249..285Z}
{Zhou}, J.-L. \& {Lin}, D.~N.~C. 2008, in IAU Symposium, Vol. 249, Exoplanets:
  Detection, Formation and Dynamics, ed. Y.-S. {Sun}, S.~{Ferraz-Mello}, \&
  J.-L. {Zhou}, 285--291

\end{thebibliography}

\begin{appendix}

\section{Hot-Jupiter pile-up on a linear distance scale}
\label{sec:appendix_pileup}

The histogram of the observed hot-Jupiter pile-up in Fig.~\ref{fig:exoplanets} is shown on a logarithmically scaled abscissa with constant bin width of 0.01\,AU. In this representation, bins appear wider in close-in orbits and thinner in more distant orbits. As a consequence, the proposed hot-Jupiter pile-up could simply be a binning artefact.

Figure~\ref{fig:exoplanets_lin} shows the same data as Fig.~\ref{fig:exoplanets} and again on a logarithmically scaled abscissa, but now using a logarithmic bin width as well. In this representation, the bins appear to have constant width in the plot, although the effective bin width really depends on the semimajor axis. Near the pile-up at 0.05\,AU, for example, the bin width is about 0.002\,AU, whereas near 1\,AU the bin width is roughly 0.05\,AU.

Technical details aside, the most important conclusion to be drawn from Fig.~\ref{fig:exoplanets_lin} is that the hot-Jupiter pile-up is not a binning artefact. In this representation using logarithmic bin width, the maximum of the observed hot-Jupiter distribution near 0.05\,AU is about five times as high as it is at around 0.03\,AU or 0.1\,AU.

%**********************************************
%Fig. A.1
\begin{figure}[h!]
\centering
\includegraphics[width=1\linewidth]{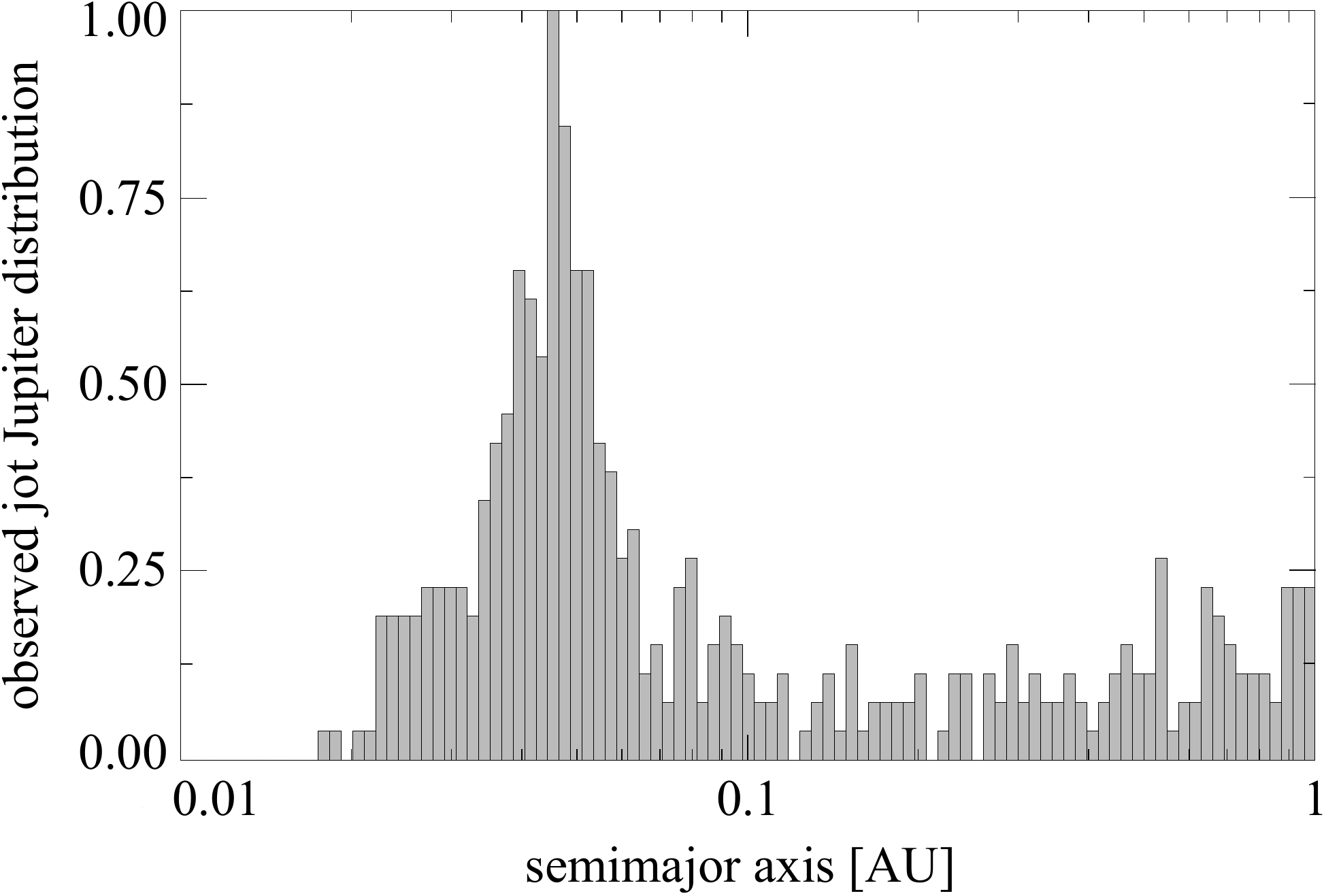}
\caption{Same as Fig.~\ref{fig:exoplanets}, but using a logarithmic bin width.}
\label{fig:exoplanets_lin}
\end{figure}
%**********************************************

\section{Lognormal randomization in gnuplot}
\label{sec:appendix_gnuplot}

The Monte Carlo simulations of Sect.~\ref{sec:results} and the illustration of Fig.~\ref{fig:torque}(b) were generated from a single {\tt gnuplot} script \citep[{\tt pile-up.gp};][]{2018ascl.soft06014H} and using {\tt gnuplot} version 5.2. Both the PDF file of Fig.~\ref{fig:torque}(b) and the estimated hot-Jupiter survival rate of 31\,\% are direct outputs from this script.

The implementation of the randomized drawings in {\tt gnuplot} code deserves some explanation because to the best of our knowledge there exists no simple way in {\tt gnuplot} to sample a random variable from a given probability distribution. The aim was to sample $\Sigma_{\rm p,0}$, $\langle\mathcal{D}\rangle_\omega$, $\alpha$, and $\mu$ based on the probability distributions of each of these random variables. As an example, we consider $\mu = 1.85\,\pm0.55$, where 1.85 is the mean value and 0.55 is the standard deviation ($1\,\sigma$) of a normal distribution around the mean. The symmetric interval of $\pm\,1\,\sigma$ around the mean  contains about 68.27\,\% of all realizations for a large number of samples.

Although {\tt gnuplot} does not have a built-in function to directly sample a probability distribution, it does have a built-in function {\tt rand(0)} to generate a random real number within [0,1] with constant probability density throughout the interval. We can combine {\tt rand(0)} with the built-in function {\tt invnorm()}, which is the inverse function of the cumulative normal distribution

\begin{equation}
{\tt norm(y)} \equiv  \int_{-\infty}^{y} {\rm d}x \ \frac{1}{\sqrt{2\pi}} e^{-x^2/2} \ ,
\end{equation}

\noindent
for our purpose. {\tt invnorm()} is defined within [0,1] and in particular we have {\tt invnorm( norm(x) ) = x}. The operation {\tt y~=~invnorm(~rand(0)~)} is then equivalent to a random sampling of values of {\tt y} according to a probability density that is given by a normal distribution.

In Fig.~\ref{fig:invnorm}, we show {\tt invnorm(x)}. We note that {\tt invnorm(0.5$~\pm~$1$\sigma$/2)}$~={\pm}1$, $\lim_{x\,\rightarrow\,0}({\tt invnorm(x)})~=~-\infty$, and $\lim_{x\,\rightarrow\,1}({\tt invnorm(x)})~=~+\infty$. The $1\sigma$ confidence interval extends from $x=0.5-1\sigma~{\approx}~0.159$ to $x=0.5-1\sigma~{\approx}~0.841$ on the abscissa and from $y=-1$ to $y=+1$ on the ordinate. As a consequence, a large number of randomized realizations {\tt y~=~invnorm(~rand(0)~)} will produce a normal distribution of {\tt y} with a mean of zero and a standard deviation of 1. We can scale the width of the standard deviation by multiplication of {\tt invnorm(~rand(0)~)} with the desired $1\sigma$ value. As an example, the {\tt gnuplot} implementation of our randomized drawings of $\mu = 1.85\,\pm0.55$ reads 

\begin{verbatim}
W = 0.55 * invnorm( rand(0) )
mu_RAND = (1.85+W),\end{verbatim}

\noindent
where the temporary variable {\tt W} is one particular realization from the normal distribution.

%**********************************************
%Fig. A.2
\begin{figure}[h!]
\centering
\includegraphics[width=1\linewidth]{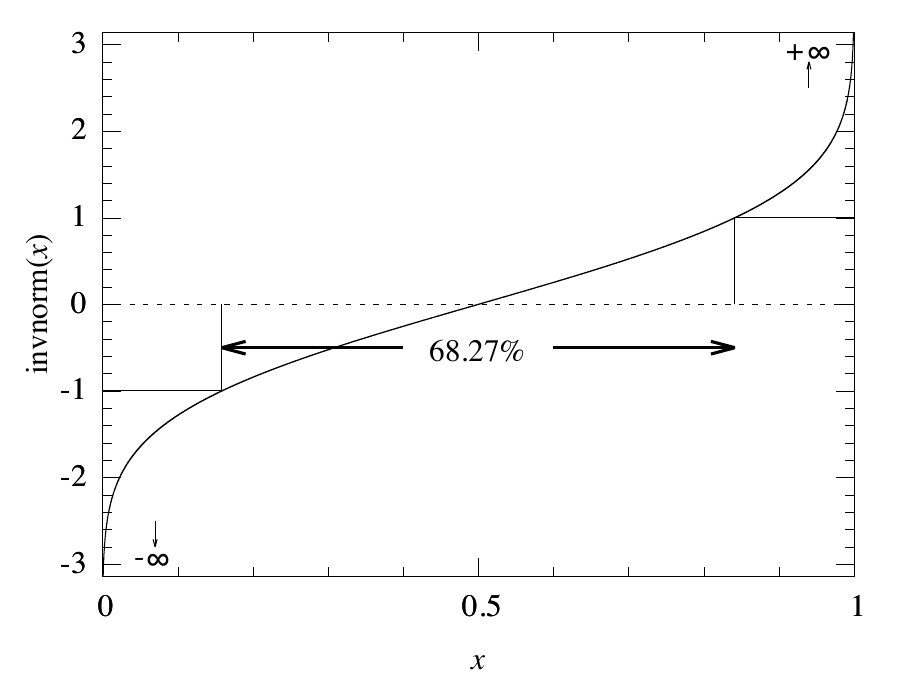}
\caption{The solid curve shows the built-in {\tt gnuplot} function {\tt invnorm(x)} that we used to generate Monte Carlo simulations of our star--planet--disk model.}
\label{fig:invnorm}
\end{figure}
%**********************************************

\end{appendix}

\end{document}